%% file: main.tex
\PassOptionsToPackage{table,xcdraw}{xcolor}

\documentclass[sigconf, nonacm, authorsperrow=4]{acmart}

\AtBeginDocument{%
  \providecommand\BibTeX{{%
    \normalfont B\kern-0.5em{\scshape i\kern-0.25em b}\kern-0.8em\TeX}}}

\makeatletter
\def\@ACM@authorsperrow{4}
\makeatother

\usepackage{appendix}
\usepackage{soul} 

\usepackage{subcaption}
\usepackage{stfloats}

\usepackage{graphicx}
\usepackage{multirow}
\usepackage{pifont}
\usepackage{lscape}

\usepackage{mathtools}

\renewenvironment{quote}
  {\list{}{\leftmargin=2em\rightmargin=2em}\item\relax}
  {\endlist}

\RequirePackage[normalem]{ulem} 
\RequirePackage{color}\definecolor{RED}{rgb}{1,0,0}\definecolor{BLUE}{rgb}{0,0,1} 

\begin{document}

\title[Unseen City Canvases: Exploring BLV People’s Perspectives on Urban and Public Art Accessibility]{Unseen City Canvases: Exploring Blind and Low Vision People’s Perspectives on Urban and Public Art Accessibility}

\author{Lucy Jiang}
\affiliation{%
  \institution{University of Washington}
  \city{Seattle, WA}
  \country{USA}}
\email{lucjia@uw.edu}

\author{Amy Seunghyun Lee}
\affiliation{%
  \institution{University of Washington}
  \city{Seattle, WA}
  \country{USA}}
\email{slee45@uw.edu}

\author{Jon E. Froehlich}
\affiliation{%
  \institution{University of Washington}
  \city{Seattle, WA}
  \country{USA}}
\email{jonf@cs.uw.edu}

\author{Leah Findlater}
\affiliation{%
  \institution{University of Washington}
  \city{Seattle, WA}
  \country{USA}}
\email{leahkf@uw.edu}

\renewcommand{\shortauthors}{Jiang et al.}

\begin{abstract}
\input{sections/abstract}
\end{abstract}





\begin{teaserfigure}
    \centering
    \includegraphics[width=\textwidth]{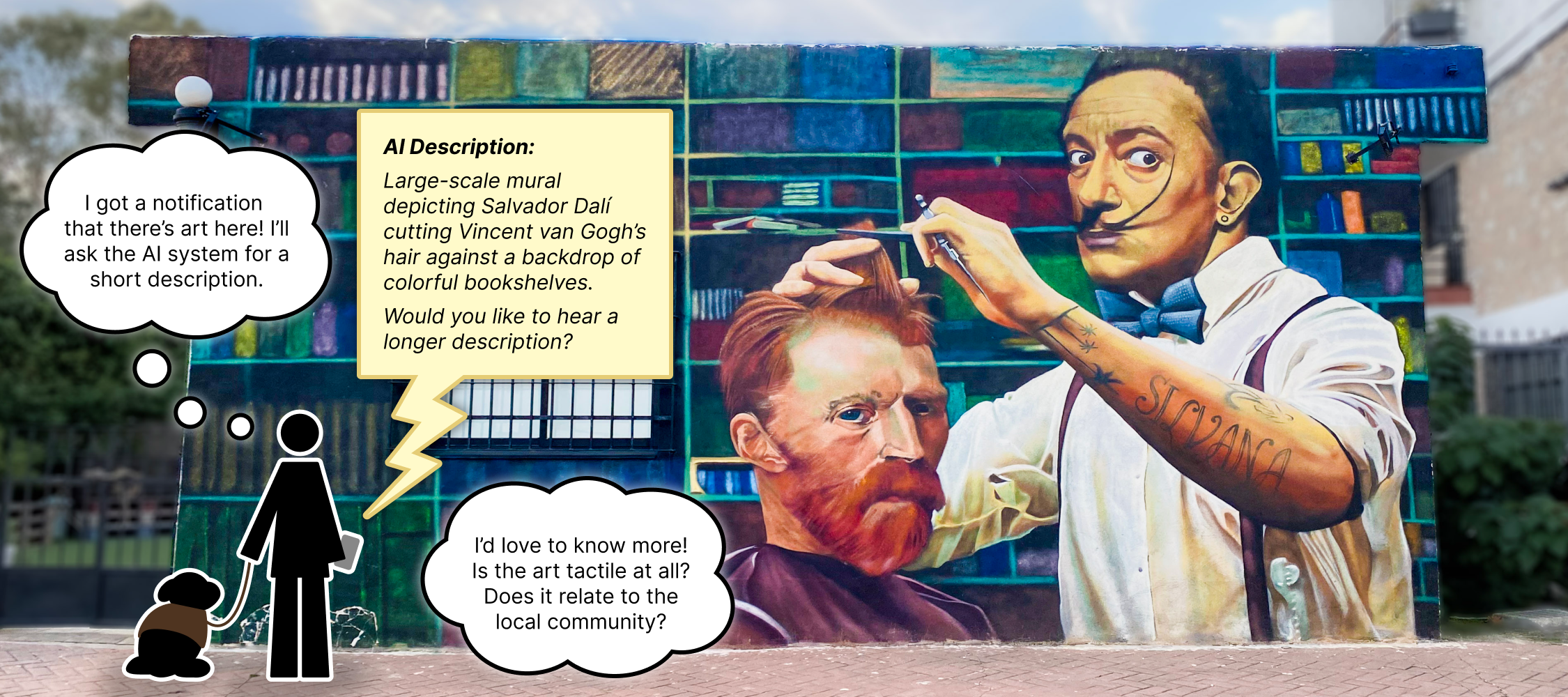}
    \caption{An example of a potential public art encounter. In this case, a blind person with a guide dog receives a notification of nearby art, listens to an AI-generated description, and asks for further detail. (\textit{Untitled} - Nesui)}
    \Description{A silhouette of a cartoon person and guide dog in front of a large, colorful mural covering an entire building facade. A thought bubble on the person’s left reads “I got a notification that there’s art here! I’ll ask the AI system for a short description.” From their phone, which they hold in their right hand, a large yellow text bubble reads “AI Description: Large-scale mural depicting Salvador Dalí cutting Vincent van Gogh’s hair against a backdrop of colorful bookshelves. Would you like to hear a longer description?” Beneath this text bubble is another thought bubble from the person, reading “I’d love to know more! Is the art tactile at all? Does it relate to the local community?”}
    \vspace{0.4em}
  \label{fig:teaser}
\end{teaserfigure}

\maketitle
\input{sections/initial}

\bibliographystyle{ACM-Reference-Format}
\bibliography{references}

\newpage
\onecolumn
\appendix
\input{tables/proberationale}
\newpage
\input{tables/probedescriptions}

\end{document}

%% file: sections/abstract.tex
Public art can hold cultural, social, political, and aesthetic significance, enriching urban environments and promoting well-being. However, a majority of urban art is inaccessible to blind and low vision (BLV) people. Most art access research has focused on private and curated settings (e.g., museums, galleries) and most urban access work has centered on outdoor navigation, leaving urban and public art accessibility largely understudied. We conducted semi-structured interviews with 16 BLV participants, using design probes featuring AI-generated descriptions and real-time AI interactions to investigate preferences for both discovering and engaging with urban art. We found that BLV people valued spontaneous art exploration, multisensory (e.g., tactile, auditory, olfactory) engagement, and detailed descriptions of culturally significant artwork. Participants also highlighted challenges distinct to urban art contexts: safety took precedence over art exploration, multisensory access measures could be disruptive to others in the public space, and inaccurate AI descriptions could lead to cultural erasure. Our contributions include empirical insights on BLV preferences for urban art discovery and engagement, seven design dimensions for public art access solutions, and implications for expanding HCI urban accessibility research beyond navigation.

%% file: sections/initial.tex
\section{Introduction}
Public art is critical to shaping urban environments, fostering community identity, enhancing aesthetic appeal, and improving economic outcomes~\cite{sharp2020just, schuermans2012public, cheung2021impacts, knight2008public, kuhnapfel2025impact, cartes1997art}. Public and urban artworks\footnote{Throughout this paper, we use the terms \textit{“public art”} and \textit{“urban art”} interchangeably to refer to artworks in publicly accessible outdoor urban spaces.} are often community-driven, rooted in activism, and directly connected to their audience~\cite{knight2008public, lacy1995mapping, zebracki2013beyond}. However, public art is rarely accessible to blind and low vision (BLV) people due to barriers in both discovering and engaging with the artwork~\cite{vocaleye, vocaleyecrowder, kleege2017more, gleason2018footnotes}, excluding them from its significant aesthetic, cultural, and community-building benefits. 

Prior work by both practitioners and researchers has examined the accessibility of visual arts in museums, galleries, and other private settings. For example, some museums provide descriptive touch tours, verbal descriptions, and braille and digital versions of print information~(e.g.,~\cite{smithsoniantouchtour, cooperhewittguidelines, metresources}). Researchers have explored integrating touch, audio, and description capabilities for multisensory art experiences~\cite{rector2017eyes, cavazos2021accessible, holloway2019making} as well as leveraging AI to generate artwork descriptions~\cite{chheda2024engaging, chheda2025artinsight, copilotart}. Most prior work on art access is limited to museums and galleries, which are highly curated contexts featuring defined pathways and centralized organization~\cite{jiang2024making}. Furthermore, prior work on BLV urban accessibility has focused primarily on functional applications, such as navigation and wayfinding~(e.g.,~\cite{jain2024streetnav, gleason2018footnotes}), but much less research has addressed leisurely and cultural experiences in urban spaces, a right for all as affirmed by the United Nations~\cite{unarticle30}. As such, urban art accessibility is largely unexplored, leaving a gap in understanding BLV people’s preferences for public art accessibility \textemdash{} including both the exploration and engagement experiences \textemdash{} given the more dynamic, decentralized, and less organized nature of urban spaces~\cite{saha2021urban} compared to museum and gallery settings. 

In this paper, we explore two research questions: \textbf{How do blind and low vision people wish to \textit{discover} and \textit{engage with} urban and public art? How might existing art or urban access techniques apply, and what \textit{new considerations are needed for urban art contexts}?}~(e.g.,~\autoref{fig:teaser}). We conducted two-part semi-structured interviews with 16 BLV participants. First, we asked participants about their prior experiences and preferences for art in public and private settings. Then, to support speculation and ground discussion, we constructed design probes by combining diverse public artworks with AI-generated descriptions and real-time AI chat interactions. We selected artworks based on a novel seven-dimensional design space that we developed based on prior literature~\cite{knight2008public} and resources from arts organizations~\cite{projectforpublicspaces, publicartarchive, americansforthearts}. The dimensions in this design space include art attributes (\textit{type}, \textit{artist purpose}, \textit{scale}, and \textit{tactility}) and the engagement experience (\textit{visitor goal}, \textit{crowd density}, and \textit{sociability}).

From a thematic analysis, we found that BLV people’s art accessibility considerations varied between public and private art settings. Participants valued how urban and public artworks were connected to local culture, interactive, and approachable, but they faced significant discovery barriers and wished to have both spontaneous notifications and curated databases for planning visits. Participants also highlighted critical safety concerns within busy urban environments, such as overlapping audio streams from urban infrastructure, vehicles, and artwork descriptions. To improve artwork accessibility, participants wished to have multisensory enhancements such as touch, verbal descriptions, audio, and scent. For example, 3D printed or crowdsourced scaled replicas could improve access to unreachable sculptures~(e.g.,~on tall pedestals, suspended in midair), while music could support a gradual, exploratory experience with artwork in the urban space. In tandem, participants highlighted critical art exploration challenges unique to outdoor public spaces, including weather, vandalism, and disrupting others in the urban environment. While all participants were generally positive about AI-generated artwork descriptions and appreciated receiving additional detail from their follow-up AI chat interactions, they shared concerns about accuracy, particularly for cultural elements where errors led to significant misunderstandings.

In summary, we contribute (1)~empirical insights on the accessibility of both discovering and engaging with urban and public art, including perspectives on generative AI for art accessibility as elicited through our design probes, (2)~a seven-dimensional space for urban and public art design solutions, bridging urban art and HCI literature, (3)~a synthesis of how techniques and technology from prior art accessibility research transfer to public art as well as unique considerations for urban spaces, and (4)~implications for future HCI research focused on nonvisual access to urban spaces that go beyond functionality alone.

\section{Related Work}
We first present literature on urban and public art theory. Then, we situate our work within HCI research exploring how to make art and the settings they inhabit \textemdash{} museums, galleries, and other cultural institutions \textemdash{} more accessible. Lastly, we describe the current scope of research on urban and public space accessibility.

\subsection{Urban and Public Art Theory} \label{rw_urbanpublicart}
Public art shapes, and is shaped by, the urban environment. In our work, we generally consider urban and public art to be socially engaged~\cite{lacy1995mapping}, non-commercial~\cite{baldini2022street}, and integrated with its site or environment~\cite{hein2006public, lacy1995mapping}. Urban planning, architecture, and visual arts literature claims that public art contributes to economic, cultural, and community development, improving sustainability and well-being~(e.g.,~\cite{sharp2020just, schuermans2012public, cheung2021impacts, tan2024walking}). However, researchers warn against romanticizing public art’s impact on critical and radical urban change given its entanglement with gentrification~\cite{pinder2008urban, hall2001public, wright2018no}. 

Prior work has also examined the social, political, and aesthetic significance of public art as it relates to its audience. For example, Knight~\cite{knight2008public} proposed that populist urban art can have seven primary purposes: as monument, memorial, amenity, installation within a park, the park itself, agora, and pilgrimage. These purposes reinforce and encourage proactive, rather than passive, audience interactions; viewers have both the right and responsibility of discerning the meaning and relevance of the art~\cite{knight2008public, zebracki2013beyond}. In comparison to artwork in private institutions such as museums and galleries, public artwork should not only be economically accessible but emotionally and intellectually accessible as well~\cite{knight2008public, hein1996public, molnar2017street}. In this paper, we investigate nonvisual accessibility barriers and BLV people’s desires for exploration and engagement with public art, addressing how technology can support art access to ensure greater agency and inclusion in the social, political, and aesthetic benefits of urban art.

\subsection{Art Accessibility}
Despite efforts to improve visual art access, many barriers remain, such as a lack of description or tactile support~\cite{sullivan2020accessibility, lizhang2023understanding, bieber2013mind}. To address these challenges, HCI researchers have leveraged technology for multimodal content translation~(e.g.,~tactile, auditory, olfactory) and streamlining visual description creation.

\subsubsection{Multimodal Content Translation}
Researchers have investigated how to use different modalities, such as digital interactions, tactile graphics, or auditory cues, to increase nonvisual access to art and images. For example, augmented reality enabled zooming in or viewing object contours with higher contrast~\cite{ahmetovic2021musa, goddard2024seeing}. Tactile components can also help BLV people interpret artwork more independently and interactively. Some have used braille displays and 2D tactile graphics for spatial and color representation~\cite{kardoulias2003guidelines, garcia2024access, gyoshev2018exploiting, shin2020please, cho2021tactile}, 2.5D reliefs to convey depth and spatial layouts~\cite{cavazos2021accessible, reichinger2011high, reichinger2018pictures}, and 3D models for enhanced fidelity~\cite{holloway2019making, neumuller20143d, leporini2020design, bruns2023touch}. Researchers have also explored sonification and other auditory techniques, often simultaneously integrating multiple senses~(e.g.,~\cite{luo2023wesee, nadri2019preliminary, rector2017eyes, cavazos2021multi, carlucci2025can}). To improve agency, Rector et al.~\cite{rector2017eyes} developed proxemic audio interfaces using background music, sonification, sound effects, and descriptions. Additionally, Cavazos Quero et al.~\cite{cavazos2021multi, cavazos2018interactive} used sounds, scents, and tactile sensations to convey spatial and semantic information, which offered layered detail but introduced sensory overload for some. 

In practice, there are many ongoing efforts to make art museums more tactile. Museums such as the \textit{Metropolitan Museum of Art}~\cite{mettouchtour} and the \textit{Art Institute of Chicago}~\cite{chicagotouchtour} offer multimodal art engagement through descriptive touch tours, and some institutions provide braille and digital versions of print information~\cite{smithsonianresources, metresources}. At the exhibition level, curated experiences such as the \textit{Tate Sensorium}~\cite{pursey2018tate} were not conceptualized for accessibility but integrated touch, taste, sound, and smell into a visual art exhibition. Lastly, some grassroots art accessibility initiatives~(e.g.,~\cite{unseenart, blindposse, vocaleyecrowder}) create 3D models and audio description to expand visual art access beyond the limits of a single museum or site.

\subsubsection{Visual Descriptions}
Many BLV people use visual descriptions to access artwork. Description creators often refer to \textit{Art Beyond Sight} guidelines~\cite{axel2003art, artbeyondsight} as a resource in both research and practice~\cite{henrich2014case, hoyt2013emphasizing}. Guidelines suggest providing a general overview of the artwork’s subject, form, and color and following up with vivid but still objective details~\cite{axel2003art, artbeyondsight}. While artwork descriptions build on image description principles~(e.g.,~\cite{stangl2020person, stangl2021going, bennett2021s}), there are key distinctions in terms of desired details and viewer goals~\cite{lizhang2023understanding, jiang2024making, kleege2017more, artbeyondsight}.

Researchers and accessibility organizations have explored both human-powered and AI-generated efforts to expand artwork description provision. Human-powered approaches often leverage crowdsourcing, following earlier initiatives for crowdsourced image description~(e.g.,~\cite{bigham2010vizwiz, burton2012crowdsourcing}). For example, crowdsourced and scaffolded art descriptions written by non-experts were still valuable for improving access~\cite{kwon2022supporting, corbett2021designing}. \textit{VocalEye}, a non-profit organization based in Vancouver, Canada, has also organized “Crowder” events to engage the public in describing urban murals~\cite{vocaleye, vocaleyecrowder, jiang2024artivism}. On the other hand, with rapid advancements in generative AI, recent work has investigated how large language models can be used to describe artwork. Bennett et al.~\cite{bennett2024painting} found that some BLV artists leveraged AI tools to describe and \textit{“check”} their own artwork. Others identified that descriptions generated by current models~(e.g.,~BLIP-2, GPT-4) did not provide sufficiently detailed information about key elements, spatial relationships, and artwork contexts~\cite{doore2024images}. Furthermore, Chheda-Kothary et al.~\cite{chheda2025artinsight} developed an AI-powered system to encourage context-aware engagement with child-created artwork, finding that creative descriptions improved understanding but could be perceived as too stylized.

While prior work has primarily focused on museum and gallery art accessibility, much less is known about how to address the unique challenges of urban art accessibility, such as how to integrate technological multisensory methods into dynamic public spaces and how to best describe community-centered, contextual artwork.

\subsection{Museum and Gallery Accessibility}
In addition to making the art itself more accessible, HCI researchers have investigated how to make art institutions more accessible to BLV patrons~(e.g.,~\cite{butler2023gallery, vaz2020perspectives}). From a longitudinal study, Butler et al.~\cite{butler2023gallery} proposed a framework for inclusive and accessible gallery experiences. Their framework highlighted considerations such as \textit{resources} (time, cost, technology, and expertise) and \textit{interpretation} of artwork beyond the individual artifact and in the context of the larger collection. In our work, we draw on this framework and investigate how it extends to urban and public settings. 

Prior work studied how technological interventions could support BLV people in interpreting and navigating art spaces~(e.g.,~\cite{holloway2019making, luo2023wesee, butler2023gallery, nagassa2024push, doore2019natural, butler2019closer}). Holloway et al.~\cite{holloway2019making} conducted a pilot study with a regional art gallery featuring multisensory adaptations of the artwork, which BLV patrons felt improved art accessibility, autonomy, and social connection. Others identified museum curators’ concerns with accessible technologies in exhibitions, including navigation efficacy in crowded environments, the lack of personalized wayfinding, and limited guidance for creating descriptions~\cite{huang2023understanding}. In studying science museum accessibility, Wang et al.~\cite{wang2025engaging} addressed the highly social nature of museum enjoyment, advocating for both individual autonomy and collaborative museum experiences. However, prior studies are largely limited to lab and gallery environments, rather than dynamic urban spaces.

Researchers have also studied how technology can support BLV navigation within museums. Prior work investigated wayfinding techniques using turn-by-turn directions, spatial audio, and vibration instructions~\cite{jain2014pilot, asakawa2019independent, wang2024direct}, tactile guiding indicators~\cite{meliones2018blind, chick2017co}, robot assistance~\cite{kayukawa2023enhancing, guerreiro2019cabot}, and 2D and 3D accessible maps~\cite{holloway2018accessible, wang2022bentomuseum}. However, as acknowledged by prior work~\cite{meliones2018blind, kayukawa2023enhancing}, safety is critical. Though indoor and outdoor art settings differ significantly in curation and organization, little work has studied accessible navigation for public art.

\subsection{Urban and Public Space Accessibility}
\subsubsection{Navigation and Wayfinding}
Most BLV urban accessibility research focuses on navigation and wayfinding, a critical accessibility need. Some researchers have also advocated for technology to complement BLV people’s orientation and mobility training~\cite{williams2014just, kameswaran2020understanding}, guiding this area of navigation accessibility work. Researchers have focused on helping people navigate in real-time, which includes advancing indoor navigation techniques~(e.g.,~\cite{ahmetovic2016navcog, mascetti2025navgraph}), understanding what obstacles should be described to people in the urban space~(e.g.,~\cite{hoogsteen2022beyond}), and improving navigation precision~(e.g.,~\cite{saha2019closing, jain2024streetnav}). For example, Jain et al.~\cite{jain2024streetnav} leveraged existing street camera infrastructure for real-time navigation, finding that computer vision methods were more precise than GPS. 

Some research has also investigated how to support BLV planning and exploration processes. Researchers have built systems to make street view imagery more accessible to BLV users~\cite{jain2025scenescout, froehlich2025streetviewai} and others developed technologies to help BLV people preview routes and build mental maps~\cite{guerreiro2017virtual}. This work often leverages points of interest (POIs) for navigation (i.e., identifying where to turn, being destinations in and of themselves). From a navigational lens, public artworks and other leisure-oriented POIs can serve as reference points~\cite{guerreiro2017virtual, kamikubo2024we}. However, currently identified POIs are usually limited to obstacles, infrastructure, or basic descriptions of businesses~\cite{hoogsteen2022beyond}. While important, this research focuses on moving from one point to another accurately and safely, instead of understanding BLV people’s needs for engaging with their surroundings beyond surface-level awareness. 

This functional focus leaves a gap in understanding how technology can support BLV people’s access to the cultural, social, and political dimensions of urban spaces, motivations already documented for museum and gallery artwork~\cite{lizhang2023understanding}. While museums and galleries have established frameworks for nonvisual art access, the unique challenges of accessing art in urban and public environments remain unexplored. 

\subsubsection{Describing Points of Interest}
While navigation research treats POIs primarily as functional waypoints, a smaller body of research has begun to investigate POIs as worthy of description in their own right. Gleason et al.~\cite{gleason2018footnotes} developed FootNotes, a system supporting functional, visual, historical, and social annotations on points of interest. They found that functional and visual annotations were most valuable, and historical annotations could be helpful in leisure situations. Especially as their work included public artworks~(e.g.,~sculptures) in their list of POIs, this demonstrated the dual nature of urban art as both functional and aesthetic; however, they did not specifically explore how to improve the accessibility of public art and its cultural and social impacts. Other investigations into leisure activities for BLV people have focused on outdoor open spaces~(e.g.,~\cite{bandukda2020places, gupta2024sonicvista}) and window shopping~(e.g.,~\cite{kamikubo2024we, kaniwa2024chitchatguide}). For example, Gupta et al.~\cite{gupta2024sonicvista} investigated sonification for outdoor scene awareness, finding that auditory icons were preferred for leisure contexts. Similarly, Kamikubo et al.~\cite{kamikubo2024we} identified the importance of having both push and pull notifications to customize description detail for window shopping, though they focused on a structured mall context with less explicit connection to cultural, social, or political aspects of public spaces. 

Thus far, art accessibility work has explored limited settings, with little to no insights on how to best describe the historical or visual components of urban artworks. Furthermore, urban access research on navigation or describing surroundings is distinctly different from experiencing urban art and supporting cultural exploration through visual details. This leaves a gap at the intersection of these two fields: \textbf{urban art accessibility research}. In this work, we examine how technology can support BLV people’s access to the benefits of discovering and engaging with public art, bringing focus to leisure and cultural access in accessibility research.

\section{Design Space for Urban and Public Art} \label{designspace}
\begin{figure*}[h]
    \centering
    \begin{subfigure}[]{0.48\textwidth}
        \includegraphics[width=\textwidth]{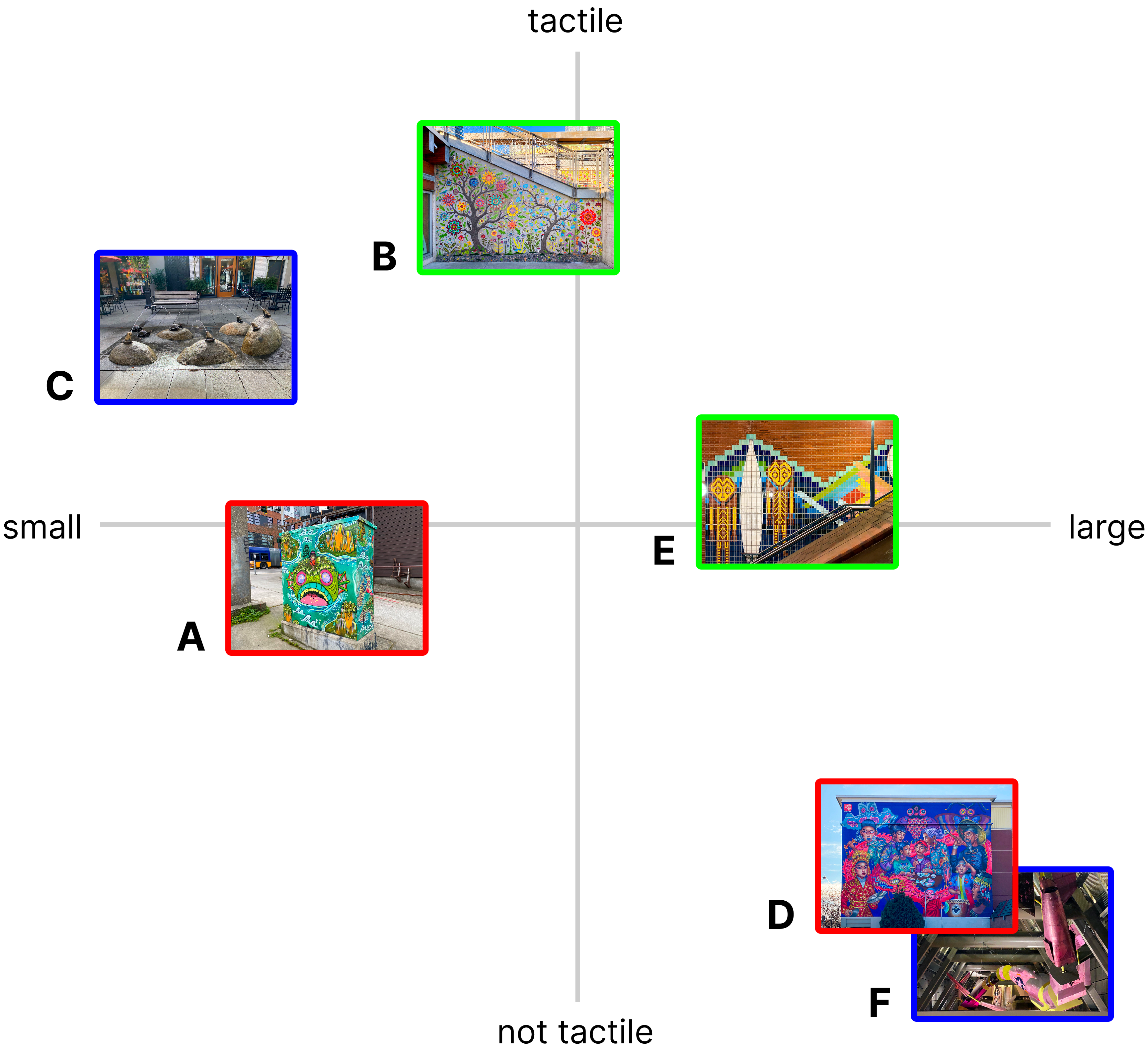}
        \caption{Plot of the artwork probes along two design dimensions: \textit{scale} (small to large) and \textit{tactility} (reachable to not).}
    	\Description{A two-dimensional graph with scale, from small to large, on the x-axis and tactility, from reachable to not reachable, on the y-axis. Six images representing the artwork probes are spread out on the plot. Clockwise starting from the top-left quadrant, which represents small and reachable artwork, there is one sculpture (very small, fairly reachable), then a mosaic (medium, very reachable) which is plotted on the y-axis between the top-left and top-right quadrants. On the x-axis between the top-right and bottom-right quadrants, there is a mosaic (fairly large, partially reachable). At the bottom right corner of the bottom-right quadrant, there is a mural and a sculpture (very large, very unreachable). Lastly, along the x-axis between the top-left and bottom-left quadrants, there is a mural (fairly small, partially reachable).}
        \label{fig:axes}
    \end{subfigure}
    \hfill
    \begin{subfigure}[]{0.48\textwidth}
        \includegraphics[width=\textwidth]{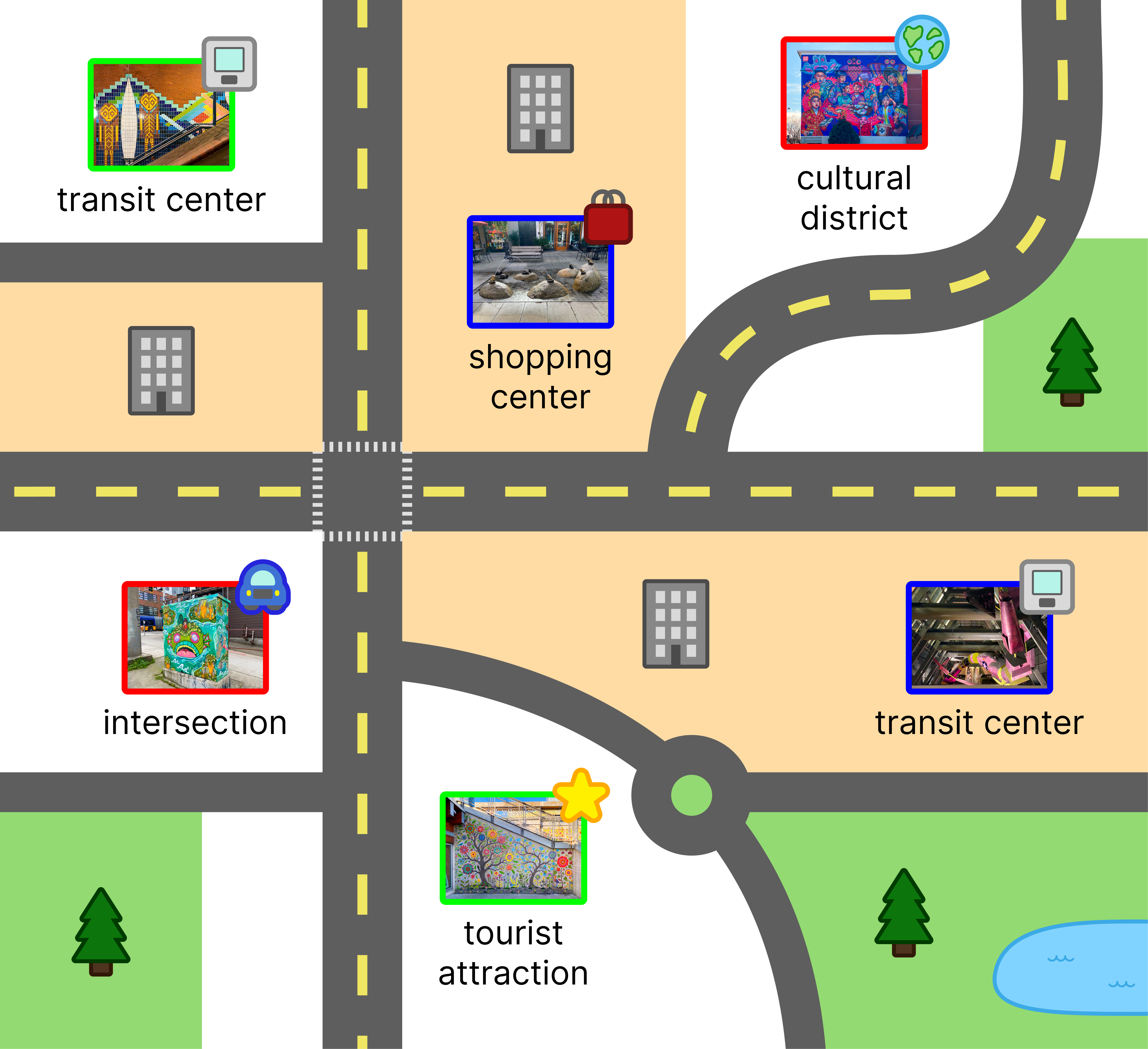}
        \caption{Map of the artwork probes in a hypothetical city, showing how \textit{visitor goal} and \textit{crowd density} may vary across locations.}
    	\Description{A cartoon map depicting the artwork probes spread out among a backdrop of large and small roads, urban areas (denoted by an orange background and building icons), and parks (denoted by a green background and tree icons). From left to right, the top half of the map shows a mosaic at a transit center (near an urban area), a sculpture in a shopping center (within an urban area), and a mural in a cultural district (near a major road). From left to right, the bottom half of the map shows a mural at a major intersection (with crosswalks), a mosaic at a tourist attraction (near a roundabout and park), and a sculpture at a transit center (within an urban area).}
        \label{fig:map}
    \end{subfigure}

    \caption{A visual illustration of our design space, drawing from~\cite{knight2008public, zebracki2013beyond, henrich2014case}. Artwork \textit{type} is indicated by color (red = mural, green = mosaic, blue = sculpture). Larger artwork images are in~\autoref{fig:probes}.}
	\label{fig:axesmap}
\end{figure*}

To formalize differences between urban and museum artwork settings, and to inform the construction of our design probes (composed of an urban artwork, an AI description, and an AI chat interaction), we drew on prior literature~\cite{knight2008public, zebracki2013beyond, henrich2014case, hein1996public} and investigated diverse artworks from public art databases~\cite{publicartarchive, americansforthearts, projectforpublicspaces}. Through careful review of prior work, such as the purposes of public artwork articulated by Knight et al.~\cite{knight2008public} and existing art accessibility approaches~(e.g.,~\cite{butler2023gallery, cavazos2021accessible}), our team defined seven dimensions. We specifically chose dimensions that differentiated public from private contexts and could have implications for the design of future accessibility solutions~(\autoref{table:designdimensions}). The dimensions span art attributes (intrinsic properties) and the engagement experience (contextual aspects of visiting urban spaces). This reflects a distinction between artwork-level features, which persist across visits, and situational factors, which vary per encounter; this also informs which accessibility solutions should be designed specifically for the artwork or adapt dynamically based on each user’s context. \autoref{fig:axesmap} demonstrates the design space with real-world examples.

\input{tables/designdimensions}

\subsection{Art Attributes}
There are four dimensions intrinsically related to the art itself: \textit{type}, \textit{artist purpose}, \textit{scale}, and \textit{tactility}. 

\textbf{Type:} the medium or modality of the art. For this paper, we focus on murals, mosaics, and sculptures. While no definitive list of public art types exists, we synthesized our list from online art repositories~(e.g.,~the Public Art Archive~\cite{publicartarchive}, Project for Public Spaces~\cite{projectforpublicspaces}). Though private art settings may also feature similar artwork types, urban artworks are generally site-specific and more integrated within the surrounding environment~\cite{hughes2009street, hein2006public}.

\textbf{Scale:} the size of the artwork from small to large. Public art manifests in a variety of sizes; the largest public art installations may exceed the size of what is possible to house within a standard museum exhibition~(e.g.,~\textit{Cloud Gate} is 33 feet high, 42 feet wide, and 66 feet long~\cite{thebean}).

\textbf{Tactility:} whether art can be physically reached and tactilely explored, ranging from fully to partially to not at all. Most museum and gallery art is not available for, or comprehensible by, touch~\cite{candlin2006dubious}. Prior work has explored how augmentative tactile components can improve accessibility of museum artwork~(e.g.,~\cite{holloway2019making, cavazos2021accessible}). In urban spaces, artwork tactility depends on placement~(e.g.,~ground level vs. elevated), natural tactility~(e.g.,~a flat mural vs. a 3D sculpture), and other physical barriers~(e.g.,~a fountain spraying water vs. a dry area)~\cite{jiang2024making}.

\textbf{Artist purpose:} the meaning and intention of the artwork. An artist’s purpose may directly inform what accessibility solutions are most appropriate. This dimension draws from Knight’s taxonomy of public art purposes~(\autoref{rw_urbanpublicart}) and the documented need to interpret art in context~\cite{butler2023gallery, artbeyondsight}. Purpose and artwork type may sometimes be correlated~(e.g.,~monuments are more likely to be sculptures) but are generally independent of each other. While some public artworks may be designed for decoration or aesthetic appeal, a core difference between public and private artwork is that many urban artworks harbor specific populist intentions~\cite{knight2008public}.

\subsection{Engagement Experience}
There are three dimensions related to one’s experience of engaging with art: \textit{visitor goal}, \textit{crowd density}, and \textit{sociability}.

\textbf{Visitor goal:} a person’s intention when visiting an urban space. Like anyone else, BLV people can visit a place for a specific cultural or artistic experience~(e.g.,~a popular tourist destination~\cite{ntakolia2022user}), pass by leisurely or serendipitously~(e.g.,~window shopping~\cite{kamikubo2024we}), or rush past while going to another location~(e.g.,~commuting through a transit center~\cite{ren2023experiments}). Conversely, museum and gallery visitors’ goals are naturally to explore and engage with artwork~\cite{cotter2022people, holloway2019making}.

\textbf{Crowd density:} the number of people in the vicinity and sharing the urban space, ranging from isolated~(e.g.,~a quiet area of a park) to crowded~(e.g.,~Times Square in New York City). HCI researchers have highlighted how crowd density, both in terms of people and vehicles, can impact BLV people’s experiences with urban navigation~(e.g.,~\cite{jain2024streetnav, saha2019closing}). While museums and galleries may also vary in density, the spectrum may be more constrained due to crowd management efforts and indoor fire safety requirements.

\textbf{Sociability:} whether someone is experiencing and engaging with the art by themselves or with others. As mentioned in prior work, community and social engagement are fundamental aspects of art experiences~\cite{butler2023gallery, henrich2014case}, and leisure is often an interdependent activity~\cite{bandukda2020places}. Sociability can influence assistive technology use~\cite{shinohara2011shadow} \textemdash{} solo visitors may rely more on technological support, whereas visitors in groups may prioritize in-person shared exploration.

\section{Methodology}
To examine BLV people’s perspectives on challenges and opportunities with urban art, we conducted semi-structured interviews consisting of a reflection on prior art experiences and a design exploration with six probes (Figure \ref{fig:probes}). This study was approved by our Institutional Review Board.

\subsection{Participants}
We recruited 16 BLV participants through mailing lists and existing connections to disability organizations, using a mix of purposive and convenience sampling to reach people across the United States with diverse public art experience. Participants were required to be at least 18 years old and comfortable communicating in English. We did not require participants to have background or interest in urban and public art. Twelve participants identified as blind, two as legally blind, and two as visually impaired~(\autoref{table:participants}). Some participants had color and light perception while others were completely blind. Ten participants identified as women, five as men, and one chose not to disclose. Seven identified as white or Caucasian, four as Hispanic or South American, three as Asian or South Asian, one as Native American, and one preferred to not disclose. Participant ages ranged from 26 to 91 years old ($M = 46.2$, $SD = 16.6$). They had various levels of engagement with urban art~(e.g.,~P7 and P13 identified as artists) and lived in different states such as Washington, Illinois, New Jersey, and Florida.

\input{tables/participants}

\subsection{Procedure}
Study sessions were composed of two parts. In Part 1, we asked participants about their prior experiences with art in both urban contexts and curated settings~(e.g.,~museums, galleries). In Part 2, we introduced six urban art design probes to contextualize and ground discussion. Sessions lasted approximately 90 minutes and were conducted remotely via Zoom. Our virtual approach enabled detailed exploration of a wider range of artwork probes and greater geographic diversity of participants. Participants were compensated with a \$45 gift card upon completing the study. We invited, but did not require, participants to share a public artwork of interest to discuss during the study. 

\textbf{Part 1: Prior Art Experiences.} We first asked participants about their experiences with urban and public art, including how they engaged with art, where it was, how they found out about it, and their general perspectives. We then inquired about key challenges with accessing art non-visually, how they discovered art pieces, and factors that lessened engagement or interest. To contrast these experiences with curated art contexts, we asked about prior experiences at museums and galleries, how access and discovery methods might translate to public settings, and how their goals and experiences differed in urban environments.

\begin{figure*}[!ht]
    \centering
	\begin{subfigure}[]{0.32\textwidth}
        \includegraphics[width=\textwidth]{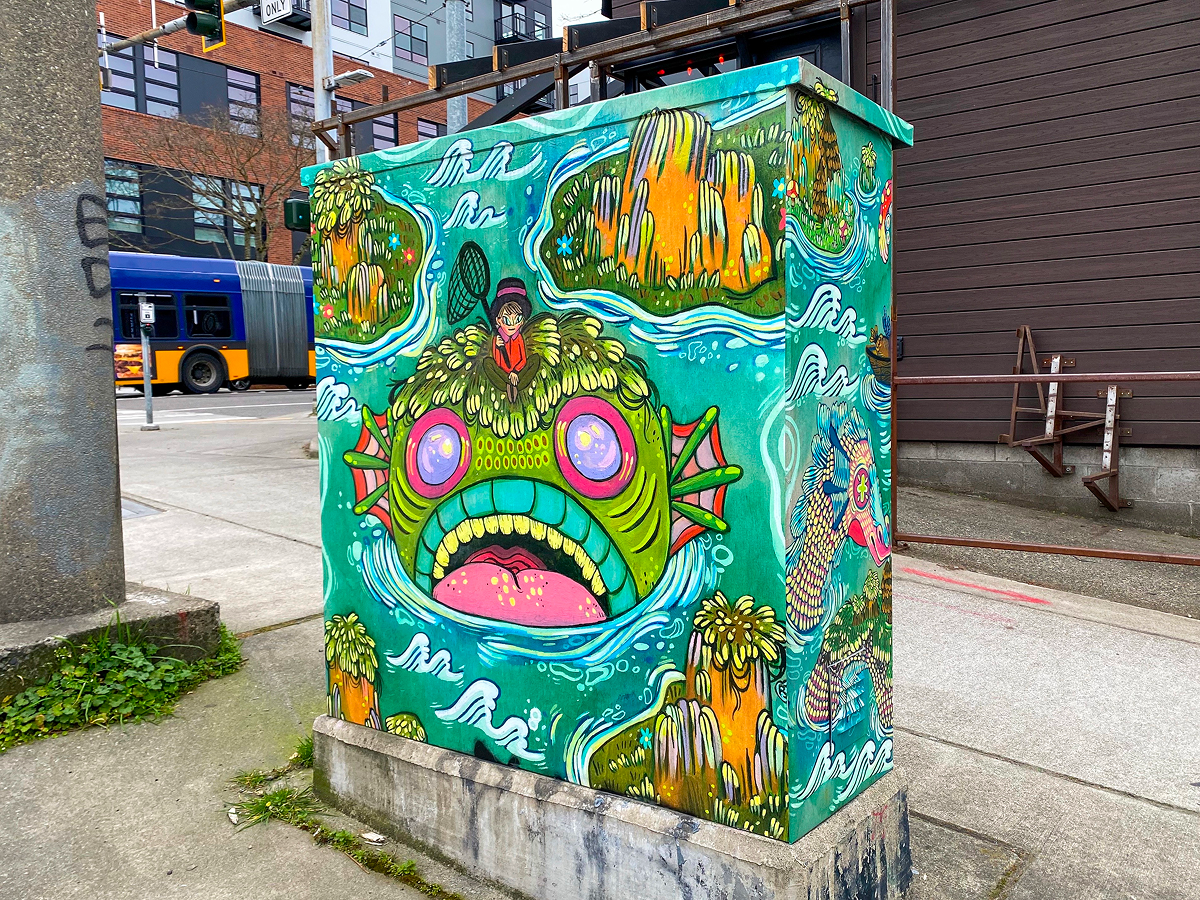}
        \caption{\textit{Untitled} - Rhodora Jacob. \\A fantastical aquatic mural on an electrical box located at an intersection.}
    	\Description{A colorful mural on a roughly five-foot tall electrical box, with the center of the main panel featuring a stylized large yellow-green sea monster with purple eyes and yellow teeth and a pink tongue in its open mouth. There is a small humanoid figure wearing a red shirt and carrying a small fishing net sitting atop the monster’s head. The sea monster is surrounded by a turquoise background with white waves and there are small orange rocky formations on grassy islands in each of the four corners of the electrical box. On the side panel, there is a pink and yellow sea dragon and additional islands and waves.}
        \label{fig:probe1}
    \end{subfigure}
    \hfill
    \begin{subfigure}[]{0.32\textwidth}
        \includegraphics[width=\textwidth]{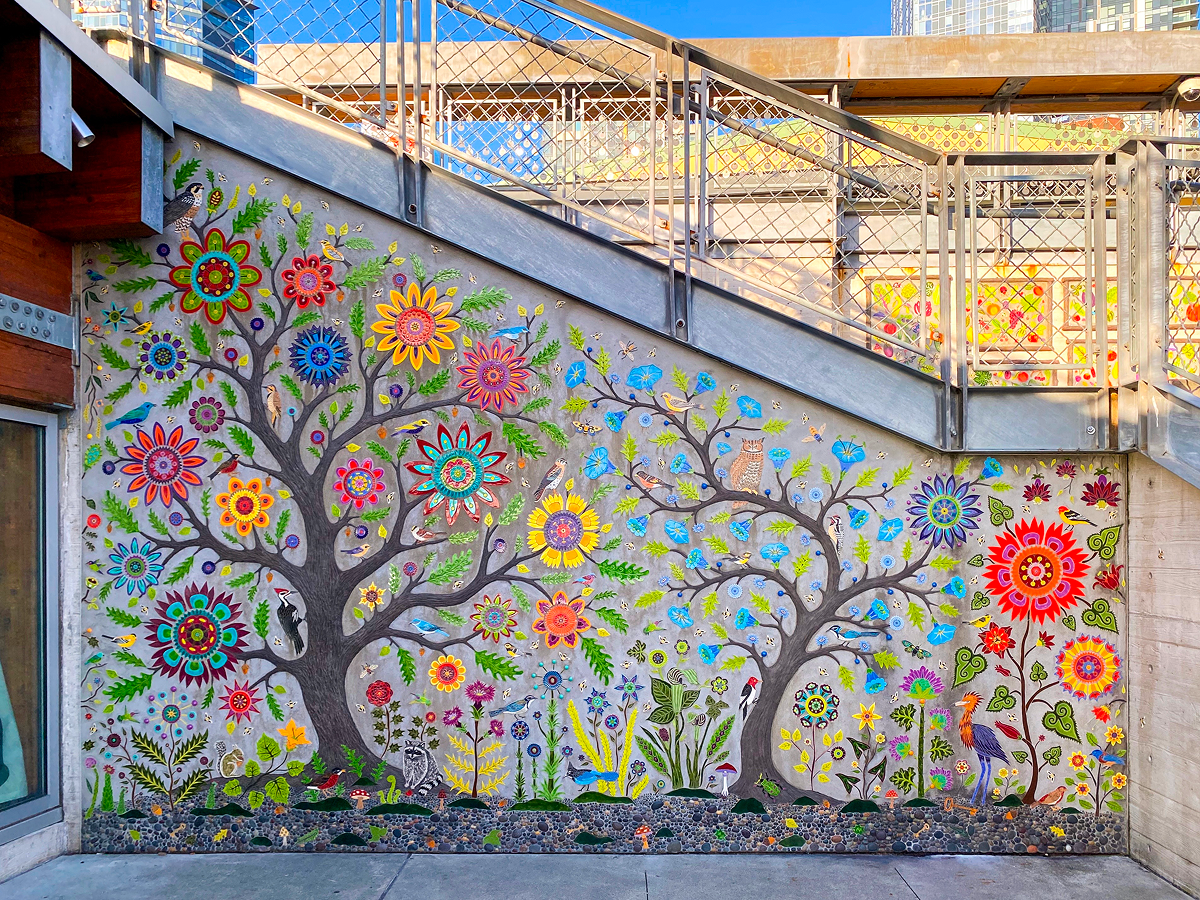}
        \caption{\textit{Northwest Microcosm} - Clare Dohna. \\A floral mosaic on a wall beneath a staircase at a tourist attraction.}
    	\Description{A mosaic featuring trees, flowers, and wildlife spanning the area beneath a staircase, measuring approximately ten feet tall by fifteen feet wide. A large tree on the left has a tactile trunk and blooms with dozens of tile flowers resembling mandalas, or concentric and symmetrical circular floral patterns, in bright reds, yellows, blues, purples, and oranges. The tree on the right primarily features light blue flowers resembling bluebells. Small birds such as woodpeckers and owls are interspersed among the branches, and other wildlife such as a squirrel and a raccoon sit on pebbles at the foot of the artwork. The colorful, glossy flower tiles contrast against the matte gray concrete in which they are inlaid.}
        \label{fig:probe2}
    \end{subfigure}
    \hfill
    \begin{subfigure}[]{0.32\textwidth}
        \includegraphics[width=\textwidth]{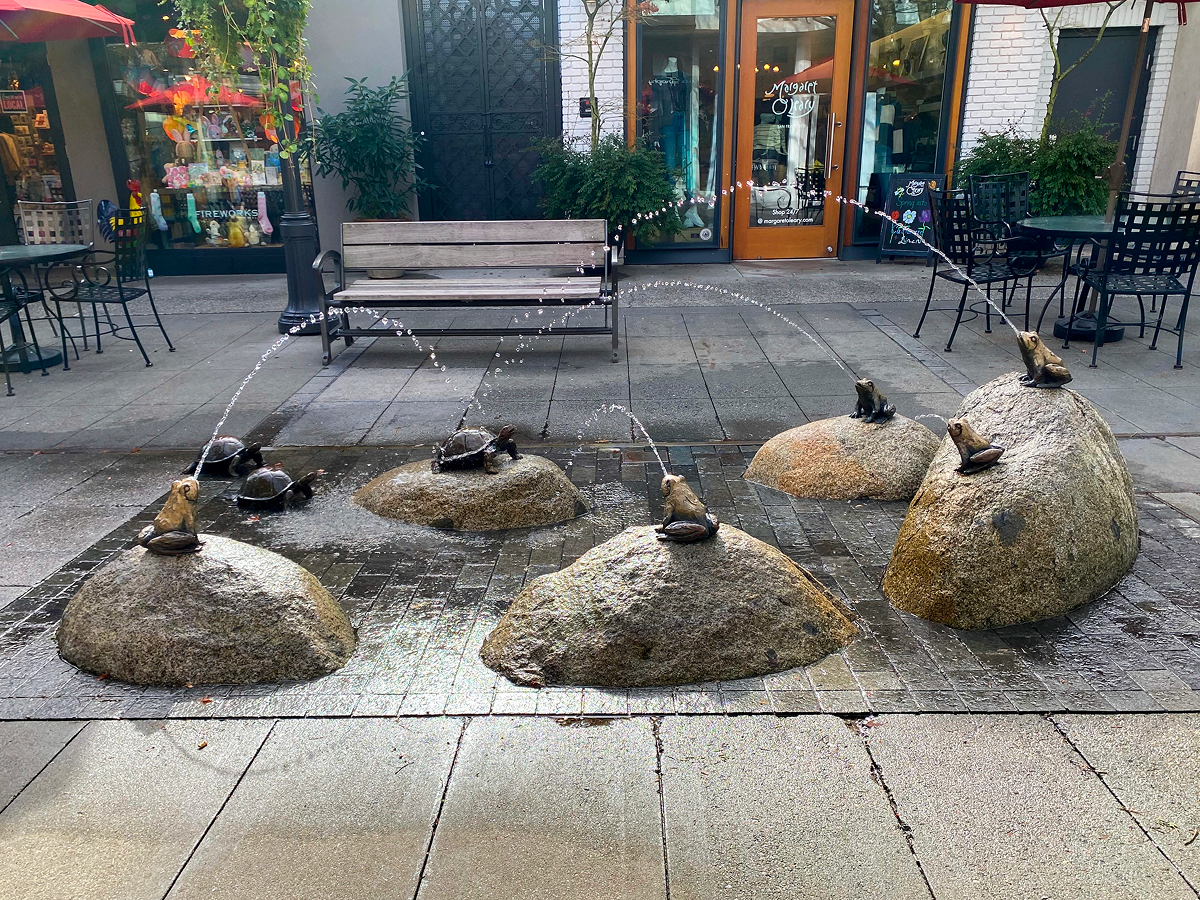}
        \caption{\textit{Water Frolic} - Georgia Gerber. \\An open fountain with mini animal sculptures in an outdoor shopping center.}
    	\Description{A rectangular sculptural fountain featuring bronze frogs spraying thin arcs of water toward bronze turtles, approximately ten feet wide by five feet long and flush with the surrounding walking area. The frogs, slightly larger than life-size, are spread out across four large, warm-toned, textured boulders in a semicircular formation, with their water streams all aiming at one turtle on another boulder. Two additional turtles are near the border of the fountain’s footprint. The immediate surrounding area is slightly wet and there are other urban fixtures such as benches, tables, and storefronts behind the fountain.}
        \label{fig:probe3}
    \end{subfigure}

    \vspace{0.25cm}
    
    \begin{subfigure}[]{0.32\textwidth}
        \includegraphics[width=\textwidth]{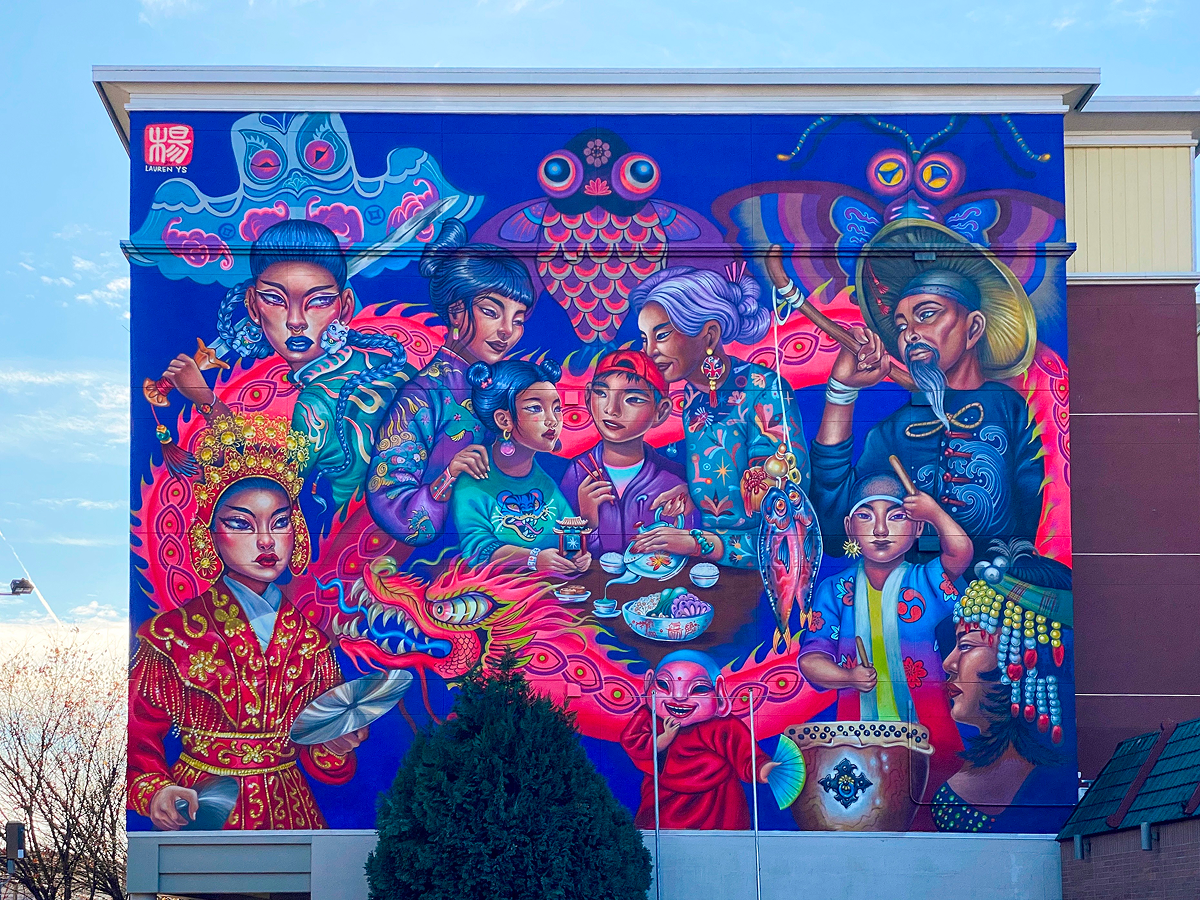}
        \caption{\textit{Untitled} - Lauren YS. \\A pan-Asian cultural mural along the wall of a building above ground level.}
    	\Description{A large, multi-story mural featuring a collage of diverse Asian figures, as well as mythical creatures such as dragons and koi fish in bright colors against a deep blue background. In the center, one adult, one older adult, and two children share tea and a meal at a table, with the adults wearing teal and purple patterned traditional garments (resembling qipao) while the children wear similarly colored modern clothing such as casual sweaters and a baseball cap. A figure in the top left wields a long sword and wears a long braid, and a figure in the bottom left wears a bright red ceremonial garment with gold accents and an ornate gold headdress, and holds silver cymbals in both hands. To the right, a bearded person wearing a straw conical hat and a dark blue robe with a wave pattern holds a wooden pole with fish hanging from it. Below is a taiko drummer wearing a multicolored robe (resembling a happi) actively drumming, a person wearing a beaded headdress, and a person wearing a red robe holding a bright pink buddha mask and a colorful paper fan. A brilliant red dragon snakes between the figures, and there are other symbolic animals near the top of the artwork. At the top left corner, there is a red square with a white traditional Chinese character (Yang) written within, with the words “Lauren YS” beneath. The mural is approximately 25 feet tall and 35 feet wide, spanning roughly the second to fourth stories of the building.}
        \label{fig:probe4}
    \end{subfigure}
    \hfill
    \begin{subfigure}[]{0.32\textwidth}
        \includegraphics[width=\textwidth]{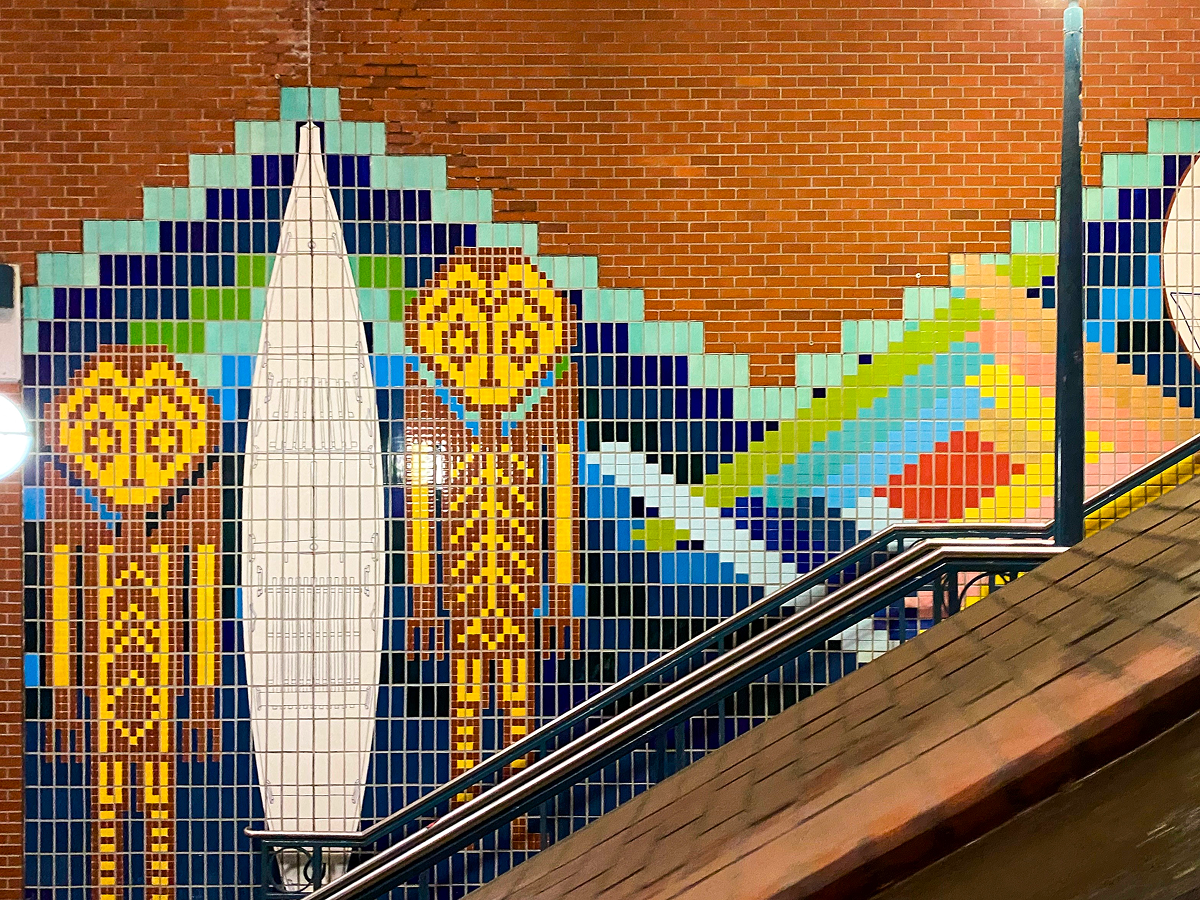}
        \caption{\textit{Untitled} - Laura Sindell. \\A mosaic with Indigenous American elements along a staircase at a transit station.}
    	\Description{A large tile mosaic in a pixellated style featuring Indigenous American imagery, approximately 20 feet tall and 30 feet wide, along a staircase and against a red brick wall. On the left, a white canoe with pale grey details stands upright between two almost identical brown and yellow totemic figures with circular heads and patterned bodies. The background of the mosaic, composed of dark blue, light blue, and green subway tiles, resembles a mountain range and ascends with the staircase. To the right of the totems there are intersecting lines of light blue, orange, yellow, and red tiles.}
        \label{fig:probe5}
    \end{subfigure}
    \hfill
    \begin{subfigure}[]{0.32\textwidth}
        \includegraphics[width=\textwidth]{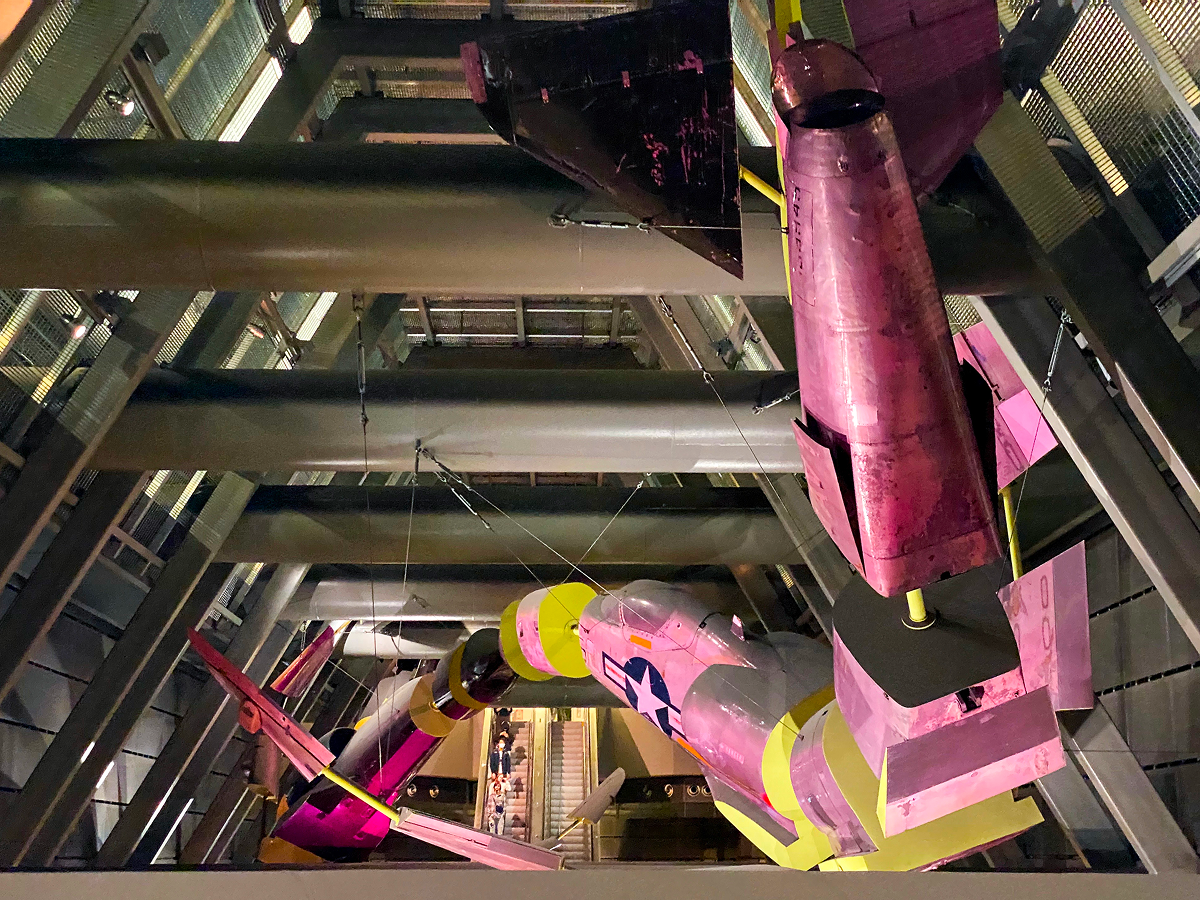}
        \caption{\textit{Jet Kiss} - Mike Ross. \\A large sculpture suspended within an open atrium space in a transit station.}
    	\Description{A large multi-piece sculpture featuring pink and yellow-painted vintage U.S. Navy jet segments, suspended from grey metal beams within an atrium. Two planes, spanning 90 feet long, are nose-to-nose. The exterior of the farther plane is painted magenta and the exterior of the closer plane is painted a lighter but still bright pink. The cross sections of the planes are painted a solid light yellow and one of the plane components has a U.S. military star emblem on its left side. The sculpture is installed far above the transit platform.}
        \label{fig:probe6}
    \end{subfigure}

    \caption{Artwork probes from the design exploration listed in the order in which they were presented to participants. Captions are paraphrased from the AI-generated short descriptions and highlight type, location, and cultural elements if relevant.}
	\label{fig:probes}
\end{figure*}

\textbf{Part 2: Design Exploration.} To further explore these ideas, we conducted a design exploration across six urban art contexts~(\autoref{fig:probes}), building on prior work using probes for data collection and participatory design~\cite{mcdonnell2021social, jiang2024context, boehner2007hci}. The probes served to extend and concretize participants’ lived experiences rather than replace them. Each probe had three components: (1)~a public artwork image representing diverse points in our design space, (2)~pre-generated short and long AI descriptions, and (3)~real-time AI chat interactions where participants could ask follow-up questions. This probe design allowed us to evaluate both static AI description quality and interactive AI dialogue for art accessibility, and facilitated exploration of different points in our design space~(\autoref{fig:axesmap}): murals, mosaics, and sculptures (\textit{type}) of different sizes and heights (\textit{scale}, \textit{tactility}) and located in diverse settings (\textit{visitor goal}, \textit{density}). For example, we categorized Probe~B as medium-sized and mostly reachable, despite its size, given the variation between the textured tiles, trunk, and concrete. In contrast, Probe~C was a fountain with small, tactile animal sculptures but the water feature and its low height made it less readily and comfortably reachable. For this study, we did not select probes based on \textit{artist purpose} as such information is not readily available for all artwork, and we could not predetermine \textit{sociability}. We return to potential design implications of these dimensions in the Discussion~(\autoref{disc_designspace}).

We used the probes as an opportunity to ground speculation and design discussions about non-visual art access, BLV art engagement and discovery, and the potential role of technology, including AI. Virtual design probes enabled us to present diverse urban artworks spanning our design space, which would have been logistically difficult to replicate through in-situ visits with participants across the United States. Prior to the interviews, we generated short and long descriptions using Claude 3.7 Sonnet\footnote{A multimodal large language model created by Anthropic, released in 2025 \cite{sonnet}.}, developing our prompt based on guidance from prior work~(e.g.,~\cite{artbeyondsight, axel2003art, lizhang2023understanding, doore2024images}) while intentionally keeping the prompt concise to approximate realistic usage of current off-the-shelf AI tools. For both description lengths, four of six descriptions were accurate; Probe~B had an artwork type error and Probe~E featured an artwork content error. Despite being prompted to do so, the short descriptions for Probe~A and Probe~C did not mention scale. We did not alter the output to improve ecological validity and to enable discussion about potential errors or ambiguities. The prompt is below and full generated descriptions are provided in~\autoref{appendix:probedescriptions}.

\begin{quote}
    “Describe this artwork with a short description that is less than 20 words. Mention what type of art it is, its scale, and other visually salient details such as the subject of the artwork and colors. Then, describe this artwork with a longer description that is around 50 words. Include details about the type of art, its scale, and other visually salient details such as the subject of the artwork and colors.”
\end{quote}

To facilitate the design exploration, we offered to share our screen to visually show the probes to all 16 participants, with seven opting in. For each probe, regardless of whether the participant was viewing the shared screen, we first read out the short description to establish a baseline understanding of the artwork unless the participant explicitly asked for the long description (\textit{N} = 1). Then, we solicited initial reactions, ideated about potential access solutions, and invited participants to ask follow-up questions to the AI system (Claude 3.7 Sonnet). The interviewer submitted each question and the corresponding probe image to the AI system, then read out the response verbatim to participants, who could ask additional questions if desired. After presenting all probes, we engaged in a reflection about preferred methods to learn about, navigate to, and engage with urban and public art. We encouraged participants to organically share suggestions before specifically introducing ideas such as spatial audio, tactile graphics, and augmented reality. We concluded with questions about the role of AI in providing artwork descriptions and the relationship between urban artwork context~(e.g.,~location, crowd density) and desired information.

\subsection{Data Analysis}
We recorded and transcribed all interviews. In line with codebook thematic analysis principles~\cite{braun2022conceptual, braun2021one}, the first author developed a preliminary codebook through hybrid coding~\cite{fereday2006demonstrating}, generating inductive codes that captured participant-driven ideas~(e.g.,~spontaneous art discovery, safety concerns) alongside deductive codes connected to theoretical concepts~(e.g.,~dimensions of the design space). The codebook was discussed with all coauthors and iteratively refined. The first author then individually coded all transcripts and the second author reviewed a randomly selected half of the coded transcripts for missed codes or unclear applications, which we resolved via discussion. We used the codebook to chart development of analysis rather than measure reliability~\cite{braun2023toward}. The final codebook contained codes pertaining to participants’ prior interactions with, perspectives on, and interest in private and public art settings. To distinguish participant-generated ideas from researcher-prompted ideas, we used distinct code prefixes for participants’ organic suggestions~(e.g.,~\textsf{suggestion: tactile: map}) and their reactions to specific technologies or designs we introduced~(e.g.,~\textsf{reaction: audio: music}). The first and second author also coded AI outputs for accuracy. For the descriptions, we adapted a subset of codes to create ternary assessments for output attributes (\textsf{accurate}, \textsf{omitted}, or \textsf{inaccurate}). For AI responses to participant questions, we identified if and how outputs contained inaccurate information (inaccuracies) or if the system declined to answer the question (refusals). With input from the research team, the first author actively constructed themes from the codes, data, and research questions to form the findings.

\subsection{Positionality}
Our team is composed of sighted researchers with backgrounds in accessibility research, including prior work in art access, tactile graphics, and AI-generated descriptions. The first author has experience as a visual arts teacher and has co-organized grassroots public art accessibility initiatives. As sighted researchers, we acknowledge that our visual engagement with artwork shaped probe selection and interpretation of participants’ experiences. To counter this bias to some degree, we spoke with a blind urban artist about practices for making his artwork more nonvisually accessible, selected artwork probes such as Probe~C~(\autoref{fig:probe3}) for their auditory, tactile, and cultural significance rather than solely visual interest, and invited participants to share and discuss urban artworks of personal interest during the study. 

\section{Findings}
We first describe BLV people’s perspectives on the similarities and differences between public and private art. Then, grounded in participants’ experiences and reactions to our design probes, we share insights on improving nonvisual exploration and awareness of urban art. We outline technological considerations for enhancing multisensory engagement with public art using tactile, auditory, and olfactory methods. We close by detailing participants’ reactions to our AI-generated design probe descriptions and general attitudes toward the potential of using AI for urban art access.

\subsection{Comparing and Contrasting Public vs. Private Art}
Unlike art in museums and galleries, which is carefully curated and catalogued in exhibits, public art is often standalone and located throughout a city. Participants particularly enjoyed how public art (1)~reflected the character of a city and motivated them to explore, (2)~was often interactive, eliciting human reactions and joy, and (3)~was generally more approachable than art in museums.

\subsubsection{Reflecting Character and Motivating Exploration}
Ten participants felt that urban art reflected a city’s vibrant character. For example, P7 described her experiences encountering public art on her daily walking commutes:

\begin{quote}
    “It gives you a sense of lightheartedness, of community, of small-town feel. [...] It doesn’t make it feel factory. It doesn’t make it look carbon copy. It gives you personality and gives you a sense of home.”
\end{quote} 

\begin{figure}[h]
    \vspace{0.2em}
    \includegraphics[width=0.75\linewidth]{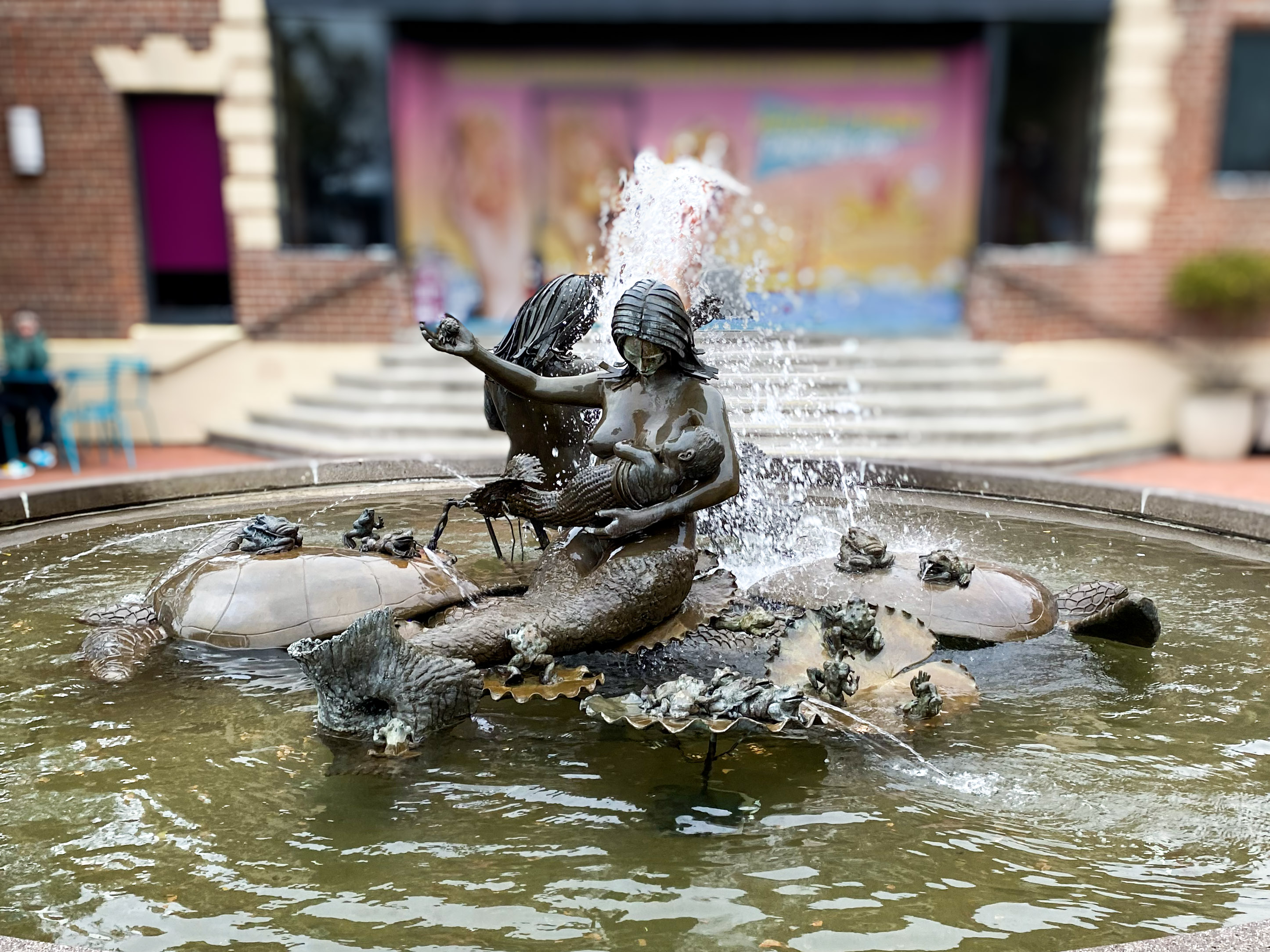}
	\caption{An intricate sculptural fountain featuring mermaids, frogs, and turtles, located in Ghirardelli Square in San Francisco, CA. (\textit{Andrea’s Fountain} - Ruth Asawa)}
	\Description{A bronze sculpture within a round fountain, featuring two mermaids sitting back-to-back above the water and two large sea turtles on the surface of the water, with small frogs atop both the turtles’ shells and nearby lily pads. The closer mermaid faces down toward the baby merperson she is nursing and she holds her right arm outstretched. Water sprays up from between the two mermaids and smaller streams flow from the mouths of the frogs.}
	\label{fig:andreasfountain}
\end{figure}

Similarly, participants mentioned how public art advanced their connection to a city’s culture, both in their own neighborhoods and those they were visiting. P12 emphasized how public art empowered local communities to have \textit{“ownership of their space,”} and P5 wished to know more about art in her local community to understand \textit{“what the artists were thinking, what their values are.”} She also cited specific artworks, given their prominence and renown~(\autoref{fig:andreasfountain}): \textit{“I would like to actually be able to fully participate. Like if someone is saying, ‘Oh, I went to Ghirardelli Square and saw Andrea’s Fountain,’ I would like to be able to have an idea in my head.”} Similarly, P10 found public art more \textit{“meaningful”} if it was \textit{“specific to the region or area that I’m visiting.”} Given its benefits, participants expressed how engaging with urban art could motivate them to explore a city, socialize, and \textit{“get out more”} (P6). 

\subsubsection{Interactivity of Public Art}
Compared to museum and gallery art, urban art more readily invited interactions from passersby. As P9 explained, \textit{“[it’s] the human side of the art \textemdash{} when it’s urban art, it’s not just the art itself, it’s the people around [and their] interaction with it.”} P16 found out about a public fountain \textit{“by accident”} and was \textit{“drawn to them just by the nature of how they sounded. [...] It’s kind of fun just to be there and listening to the water and the kids playing.”} Participants also felt the positive impact of sharing in others’ joy toward public art: \textit{“even if you’re not getting the full experience of the art, you’re experiencing it through the eyes of those who are. [...] That can be just as gratifying”} (P9).

\subsubsection{Approachability of Public Art} \label{approachability}
Participants appreciated how public art felt more approachable, less formal, and more serendipitous than museum or gallery art. P14 explained that she would \textit{“enjoy [urban art] more frequently because it is free to the public,”} and others mentioned that public art was \textit{“more of [a] surprise”} (P16). Regarding formality, multiple participants noted how urban art’s environmental boundaries were often more unconventional, creative, and interesting than art in museums:

\begin{quote}
    “When I go to a show or a gallery, it’s to truly look at the art itself. In urban settings, it’s the art within the setting. [...] Maybe that’s why I like a lot of graffiti and murals. [...] They’re using the space that’s there, and it’s not necessarily a blank, perfect canvas.” (P12)
\end{quote}

However, 12 participants emphasized safety concerns due to vehicle or pedestrian traffic, inaccessible locations, and more. They noted that the location of Probe~E, which was along a transit station stairway~(\autoref{fig:probe5}), was \textit{“prone to accident[s]”} and would not be \textit{“a convenient place [...] to be examining it for a long period. [...] Especially if it’s during rush hour, you probably get swept away with the mass of people passing through”} (P15). As a guide dog handler, P16 was also hesitant about standing still to examine urban art and \textit{“being in the way of things and having people come up and want to pet the dog”} (P16). However, he felt that \textit{“[in] a more peaceful, quiet setting, I would certainly feel more comfortable spending more time and contemplating it”} (P16).

Lastly, six participants expressed interest in having greater access to graffiti due to its deep cultural and community connection. For example, P9 advocated for accessibility to informal urban artwork types: \textit{“if there’s a part of town that is known [for] tagging and putting up really cool graffiti, I think it’s fair to give [it] [...] the same amount of artistic respect that we give to full installations.”} However, participants also acknowledged that graffiti could be controversial and some cities would not wish to \textit{“promote this behavior”} (P9) or \textit{“pay for descriptions”} (P14).

\subsection{Exploration and Awareness of Urban Art}
While engaging with art in museums is typically a planned activity, participants wished for technological methods \textemdash{} such as spontaneous notifications for exploratory discovery and databases for planned visits \textemdash{} to overcome organizational and documentation challenges unique to urban art. In either case, participants underscored that safety and awareness of surroundings was paramount.

\subsubsection{Spontaneous Notifications}
All but one participant (P7, who had some remaining vision) wished to have spontaneous notifications about nearby public art to spark their interest and enhance their cultural experience. Participants felt that spontaneous push notifications were \textit{“more fun”} (P4) and approximated the \textit{“organic”} (P9) sense of surprise that sighted people had when discovering art:

\begin{quote}
    “There’s something unique and exciting about walking through your city [...] and being able to turn a corner, and holy crud, that mural is beautiful. Or somebody did a chalk drawing that’s really amazing on the sidewalk, don’t walk on it.” (P9)
\end{quote}

Nine participants wished to use location-aware applications to receive spontaneous notifications. Some imagined that the apps could \textit{“alert me when I am [within] 10 feet”} (P2) of artwork or send \textit{“a notification that you were in front of [it]”} (P5). Despite enthusiasm for notifications in the moment, participants emphasized the importance of maintaining audible awareness of their surroundings, and prioritized navigational accuracy and safety over art exploration. For example, P4 did not want art descriptions to compromise her ability to navigate safely. P16 noted that safety concerns varied across locations, explaining how sounds in a bustling pedestrian market did not interfere with listening to descriptions, in contrast to the \textit{“nightmare”} of a street crossing: \textit{“if you’re listening to [navigation] instructions and there’s somebody else talking to you, and there’s traffic noises and there’s a light that’s beeping at you, these are all things that you really have to pay attention to”} (P16).

Aside from auditory methods, six participants suggested using tactile paving to indicate nearby urban artwork. However, they acknowledged that this was more feasible in indoor exhibits (P14) and could be expensive for \textit{“retrofitting”} (P15) existing public art. Multiple participants also highlighted concerns that additional tactile paving could interfere with its use for critical safety applications~(e.g.,~curb cuts).

\subsubsection{Pre-Planned Visits to Urban Art}
Others wished to leverage online repositories of public art containing both location information and descriptions to help with pre-planning public art visits. P12 suggested creating a \textit{“database for people with disabilities, for art in their community”} with crowdsourced descriptions. As a frequent traveler, P10 wished to know about art ahead of time \textit{“so that if I’m in the area where this public art is, I could visit it.”} Similarly, P9 thought having a map or database of urban art could help people identify specific works to seek out in person, and suggested that an app could tell a user \textit{“what’s around that we’ve already bookmarked”} upon approaching the area. For transit centers, which may already contain multiple points of interest on navigation apps, \textit{“adding the art pieces to that collection of data could be useful, because then it would all be there [in one place]”} (P16). Participants also mentioned that online repositories could support virtual engagements for anyone who could not visit artworks in person. 

\subsection{Sensory Feedback Methods: Tactile, Auditory, and Olfactory}
Participants envisioned incorporating dynamic sensory feedback that leveraged their senses of touch, hearing, and smell to improve the richness and accessibility of navigation and interpretation of urban art. For example, all 16 participants wished to use touch to interact with artwork~(e.g.,~through 3D models, tactile graphics, or directly on the art), with many preferring touch and other sensory feedback methods to form their own impressions and \textit{“notice things that [a describer] wouldn’t notice”} (P14).

\subsubsection{Tactile Challenges and Barriers}
\begin{figure}[h]
    \includegraphics[width=0.55\linewidth]{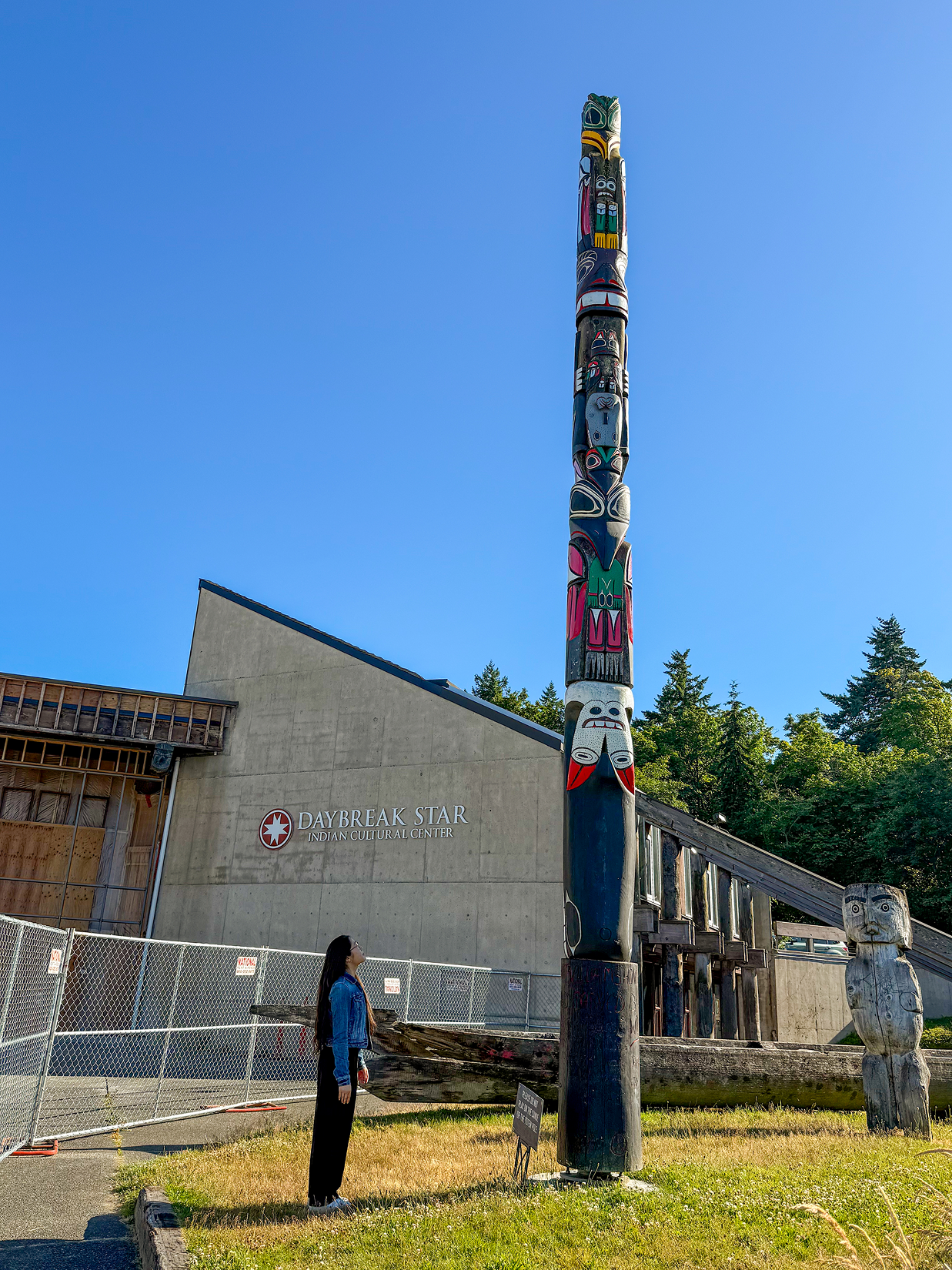}
	\caption{A visitor stands before a 34-foot tall totem pole at an Indian Cultural Center. (\textit{John T. Williams Memorial Totem Pole} - Rick L. Williams)}
	\Description{A 34-foot tall totem pole featuring an eagle, raven, and woodcarver painted with bright red, yellow, green, and white details. The visitor, blurred, is approximately five feet tall. The Daybreak Star Indian Cultural Center, a long wooden canoe, and a shorter unpainted totem pole are in the background.}
	\label{fig:totempole}
\end{figure}

Public artwork was perceived as more openly tactile than private artwork. Twelve participants explicitly expressed their frustration that museum and gallery art was \textit{“behind glass”} (P1) or enclosed from touch. Participants often used touch to understand the composition and proportions of public art, and were appreciative that \textit{“there’s nobody sitting there slapping my hand going, ‘No, don’t touch it’”} (P7). However, urban art could also be inaccessible to touch: too large to fully reach, too high above ground level, unpleasant or unsanitary, not tactile by nature~(e.g.,~murals), or located in areas that would compromise a person’s safety. For example, P4 attempted to reach a nearby sculpture but \textit{“because you’re not supposed to walk on the grass, I just stealth touched it.”} Similarly, P2 had previously encountered a large totem pole~(\autoref{fig:totempole}) but could not fully interpret its shape or details due to its size: \textit{“if I am not able to grasp [the art] with [...] two wide open arms, then I’ll have to see one chunk at a time and then put pieces together in my mind. But that can reduce [...] the whole conceptual vision of a piece.”}

\begin{figure}[!h]
    \vspace{0.2em}
    \includegraphics[width=0.75\linewidth]{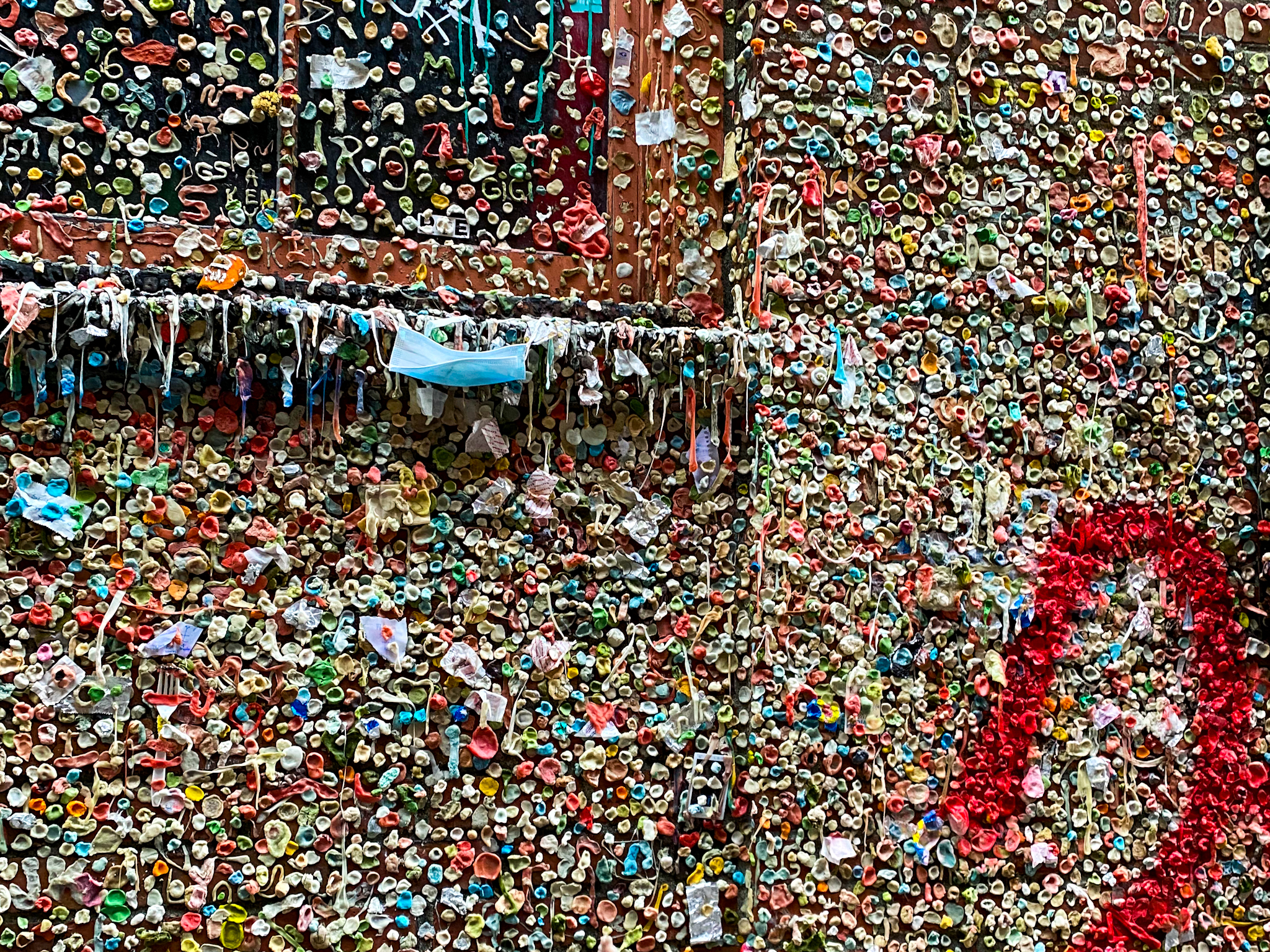}
	\caption{A wall plastered with chewed gum, which is a popular tourist attraction in Seattle, WA. (\textit{The Gum Wall})}
	\Description{Thousands of pieces of chewed gum, in all colors, densely cover the entire surface of a wall and most of a window and its sill. Most of the gum pieces are circular, while some are in creative shapes such as a smiley face or in strands as though they are melted. Other items, such as surgical masks and gum wrappers, are also stuck to the wall. A red spray paint arc covers some pieces of gum near the bottom right of the image.}
	\label{fig:gumwall}
\end{figure}

Participants also shared their experiences with public artworks that were unreachable due to being located too high up~(e.g.,~on a pedestal) or behind a barrier~(e.g.,~behind a fence). For example, participants wished to have some level of tactile engagement with sculptures because \textit{“even if you can’t reach the head, at least reach the foot. That gives you a proportion”} (P10). Additionally, while some fountains had sculptural elements, they could be impractical to reach. When discussing Probe~C~(\autoref{fig:probe3}), P14 asked the AI system if \textit{“the frogs were far enough to one side where I could actually touch it, or if they were in the dead center where I wouldn’t be able to reach it without getting completely drenched.”} Other artworks could not be interpreted via touch, either due to intrinsic tactility or sanitation risks. For example, there was \textit{“not really a whole lot for me to touch”} (P14) for murals. When describing the \textit{Gum Wall} at Seattle’s Pike Place Market, which could be considered a crowdsourced mosaic~(\autoref{fig:gumwall}), P16 mentioned: \textit{“I’m sometimes a little hesitant just because they’re just dirty, and you don’t want to drag your hands around along things that are not that clean.”} 

Finally, while most participants preferred tactile methods, some valued hands-free interactions. They emphasized how they were often already holding a cane or guide dog harness in one hand and carrying other items in the other: \textit{“we need more hands”} (P3). As a smart glasses user, P11 preferred augmented reality headsets for convenience and safety reasons: \textit{“I do not like to take out my phone when I’m in public, especially [on] trains, just for safety purposes.”}

\subsubsection{Tactile Opportunities}
Participants’ preferences for tactile urban art access largely mirrored those provided in museums and galleries, such as scaled replicas, tactile graphics, and braille displays for hands-on engagement and independent exploration. Participants mentioned that 3D models and replicas at a \textit{“a smaller scale where you could hold [it] in your hand”} (P13) could enable holistic exploration, and 11 participants suggested creating tactile graphics and maps for urban artwork. Yet the realities of public spaces introduced distinct constraints. P2 mentioned that \textit{“when it comes to urban art, there might not be necessarily a ton of money.”} While 3D printing could decrease the cost of creating tactile aids, P2 felt that in the absence of provided options, having low-cost materials (e.g., modeling clay) to create her own model was better than nothing at all. Haptic feedback, though less granular than other tactile methods, could also support interpretation of high-level information. For example, P14 suggested that vibrations could convey scale by \textit{“[indicating] the beginning and end of those big murals that are as long as a huge city building.”}

Given the variability of urban environments, participants gave suggestions for where to place tactile aids for discoverability, ease of access, and safety. Participants generally recommended keeping replicas near the original artwork, with some exceptions. At transit centers, P14 mentioned how tactile aids could be located in \textit{“waiting areas so that people could look at it while they’re waiting for the subway”}, and P4 recommended placing them near a station entrance so payment was not required for art engagement. Expensive refreshable displays~(e.g.,~a Monarch\footnote{A refreshable braille device produced by the American Printing House with a 10-line by 32-cell display, retailing for \$17,900 USD~\cite{monarch}.}) could be housed at a centralized \textit{“Monarch station”} (P6) to provide tactile access to multiple nearby public artworks. However, participants also shared concerns about damage from weather or vandalism. Living in Texas, P10 warned against using metal for braille plaques outdoors: \textit{“if it’s 100° outside and the sun is boring down, that plaque gets pretty doggone hot. [...] You can’t run your fingers over them.”}

\subsubsection{Auditory and Olfactory Feedback}
Ten participants suggested integrating music into urban artwork, especially if the art was situated within a cultural community. To make Probe~D~(\autoref{fig:probe4}) more enriching and \textit{“identifiable”} (P4), some suggested adding \textit{“Chinese traditional music[al] instruments”} (P15). Participants thought artists should be the creators of musical accompaniments, and that \textit{“the city should pay to make it multisensory”} (P14). Others shared examples of prior experiences with interactive, musical public artworks \textit{“designed with sound in mind”} (P9), which allowed visitors to contribute to the aural experience by \textit{“soundscaping”} (P9).

Participants also thought spatial audio, played either through speakers or as part of a mobile app, could support art discovery. Fourteen participants favored using spatial audio to aid with navigation to artwork, similar to familiar methods~(e.g.,~Microsoft Soundscape\footnote{An open-source spatial audio project for improving environmental awareness, originally developed by Microsoft Research~\cite{microsoftsoundscape}.}). P12 explained how auditory cues could support orienting to the intended artwork experience: \textit{“part of the art experience with installations is how the human body is poised to examine. [...] [Using] sound to guide your head to where it would be as other people are experiencing it [...] would be cool.”}

Some participants also suggested using smell for art accessibility. P10 had visited an art museum with \textit{“appropriate fragrances \textemdash{} for example, the Egyptian exhibit had an essence of frankincense.”} P3 had read about a \textit{“scented flower museum”} in New York, and highlighted how olfactory memory of perfumes and colognes helped with navigation at the mall: \textit{“it helps me know exactly where I’m at.”} 

In public settings, participants felt that spatial audio and scent could support independent, gradual, and spontaneous exploration. For example, P10 shared: \textit{“if you’re walking down the street and you suddenly smell fresh bread, [...] you’re not necessarily looking for it, but it makes itself be known to you.”} However, participants acknowledged that some access measures might not be practical or appreciated in public. Despite having positive experiences with scented museum exhibits, P16 questioned how scents could be kept \textit{“fresh”} in urban spaces. Participants were also concerned that extraneous audio could be perceived as \textit{“disruptive”} (P15) by sighted people, given existing animosity toward accessible pedestrian signals: \textit{“some people [are] already objecting to the street crossing music”} (P15). Others thought sighted people could benefit from multisensory art access: \textit{“not everybody is necessarily looking around at their environments as they’re walking. [...] But if they hear some kind of unusual music or sound, then they might look up [...] and then they would see something different”} (P10).

\subsection{AI-Generated Description Considerations}
Many participant reactions to the AI-generated description probes aligned with emerging best practices in accessible art descriptions. However, through the AI-based chat interactions during the design exploration, we uncovered preferences unique to urban art, such as cultural and historical significance, the artwork’s integration within urban space, and context-dependent information needs.

\begin{figure*}[!hb]
    \includegraphics[width=0.48\linewidth]{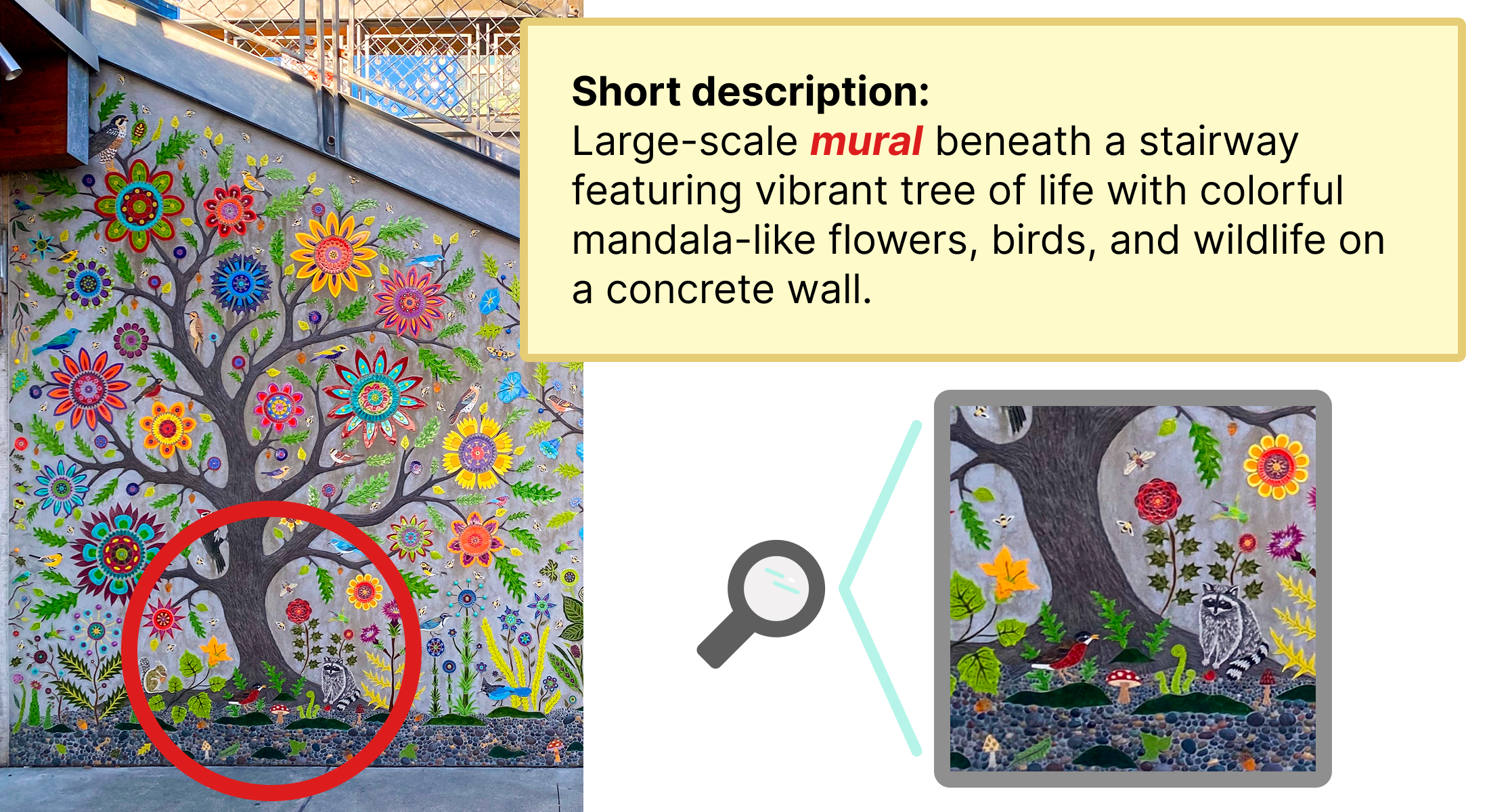}
    \hfill
	\includegraphics[width=0.48\linewidth]{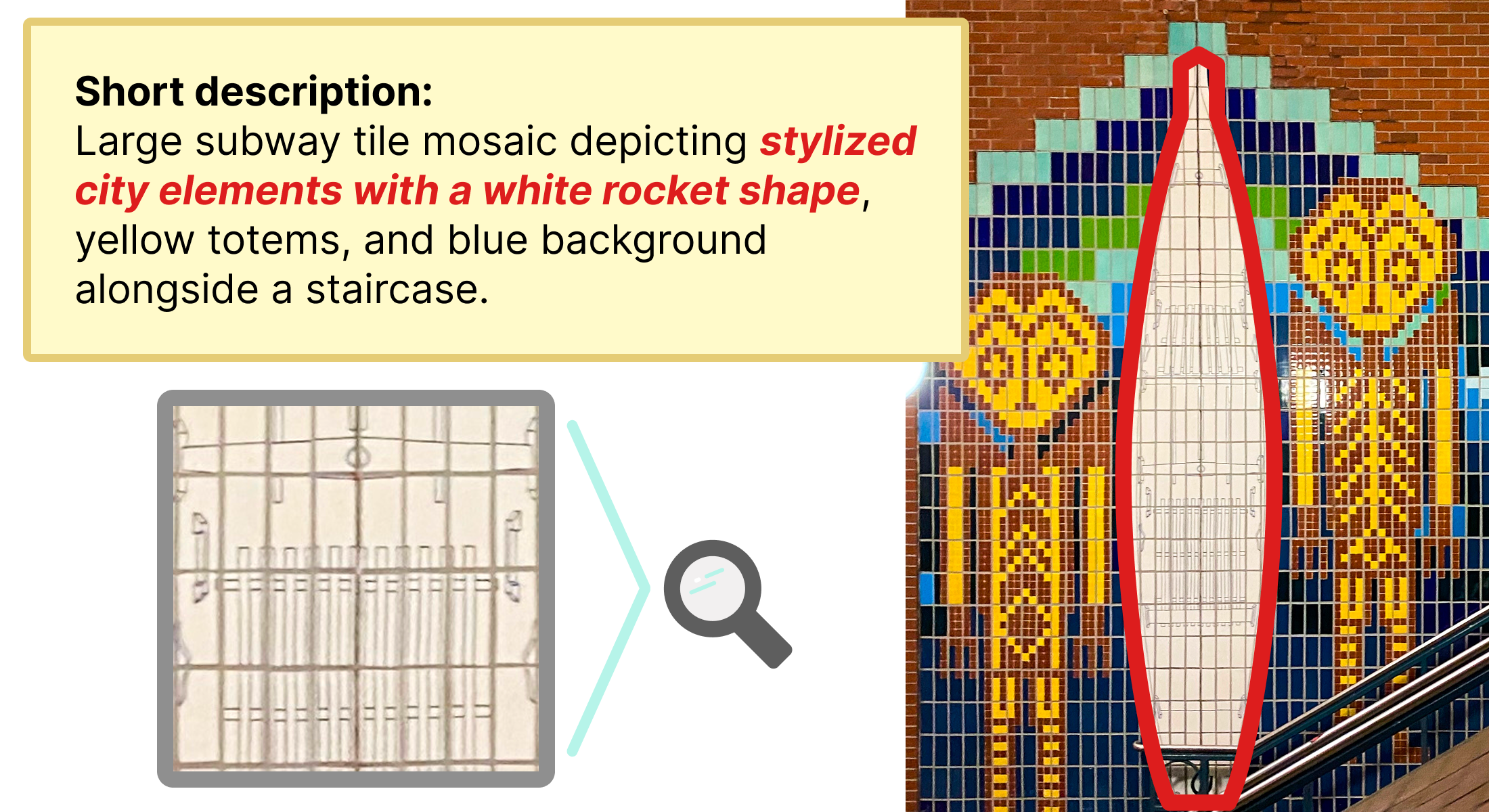}
    \caption{Close-ups of Probe~B (left) and Probe~E (right), both of which had mistakes in their AI-generated descriptions. Probe~B was characterized as a mural instead of a mosaic, and Probe~E was described as featuring a “rocket shape” rather than a canoe.}
	\Description{On the left is Probe~B, with a cropped image centered on the left half of the mosaic (one of the trees) and a red circle near the bottom of the image. A zoomed-in square shows a small raccoon sitting on the pebble-covered ground at the base of the tree trunk. The short description reads “Large-scale mural beneath a stairway featuring vibrant tree of life with colorful mandala-like flowers, birds, and wildlife on a concrete wall.” with the word “mural” emphasized. On the right is Probe~E, with a cropped image centered on the left half of the mosaic (the totems and the canoe) and a red outline around the canoe shape. A zoomed-in square shows faint grey details on the canoe such as a woven seat. The short description reads “Large subway tile mosaic depicting stylized city elements with a white rocket shape, yellow totems, and blue background alongside a staircase.” with the words “stylized city elements with a white rocket shape” emphasized.}
	\label{fig:probecloseups}
\end{figure*}

\subsubsection{Cultural and Historical Significance}
Participants valued learning more about the cultural and historical significance behind urban art, especially if it was relevant to the location or community in which the art was situated (\textit{N} = 13). When discussing Probe~D, which was located within a historical Chinatown neighborhood~(\autoref{fig:probe4}), P15 asked the AI system which Asian culture was represented to form an accurate mental image: \textit{“the colors and the dragon [...] that seems so typically Chinese [...] [but] it would have been totally different if it was Japanese or Vietnamese.”} Similarly, Probe~B featured geometric floral patterns described as “mandala-like” in the pre-generated AI short description~(\autoref{fig:probe2}). Multiple participants asked for further clarification about this visual style, with P4 noting that it was important to learn more because \textit{“they’re symbols from someone’s religion.”}

Participants also emphasized that cultural elements should be accurately and respectfully described; otherwise, this could lead to cultural erasure. For example, Probe~E~(\autoref{fig:probecloseups}) featured Indigenous American imagery. However, the AI-generated description misdescribed the canoe as a \textit{“rocket shape”} in both the short and long descriptions. Upon learning about this error, participants shared: \textit{“that totally changes it. [...] I’m envisioning, in the initial description, that this is at nighttime and there’s a rocket ship next to a city. So that’s a totally different vibe than [...] a Native American artwork”} (P14). In particular, participants felt that knowing if \textit{“Indigenous people [were] really important to this area”} could help \textit{“understand the context”} (P14) of the art, connecting back to the site-specific nature of urban and public artwork. 

\subsubsection{Environmental Integration and Boundaries of Urban Artwork}
Unlike artwork in museums and galleries, public art is typically site-specific, or designed with its location or setting in mind. P12 appreciated how \textit{“a lot of urban artists [...] look at their surroundings, and that’s different [from museums and galleries] because [...] you wouldn’t usually be designing a painting for a room in an art institute.”} Seven participants suggested including information about an artwork’s boundaries within the environment, including its spatial location, height above ground, and geographic footprint. For example, Probe~B was located beneath a staircase~(\autoref{fig:probe2}), and participants wished to know \textit{“that it’s not a rectangle [...] they made it to fit the space”} (P12). Similarly, P4 felt it would be helpful to describe the atrium containing Probe~F~(\autoref{fig:probe6}), as \textit{“it seems like it’d be a focal point if it’s suspended above the people when they walk in.”} When discussing Probe~A~(\autoref{fig:probe1}), a mural on an electrical box, P12 also wanted to know that \textit{“[the art] goes around”} and was visible from all sides. For audibly perceivable artwork such as fountains, descriptions could clarify that these water features were \textit{“a little peaceful place to sit in front of. It’s not just an obstacle [or] somebody doing road work with water dripping”} (P16).  

\subsubsection{Contextual Factors}
Description detail was dependent on a person’s relationship to the artwork location or their travel goal. For example, while some were more likely to want descriptions if the art was \textit{“somewhere I frequented, [...] if this was my [transit] stop”} (P12), others felt the opposite way: \textit{“if you use that particular station all the time, you’re just going to be passing it \textemdash{} at some point, it kind of goes into the background”} (P4). Others wanted descriptions to indicate how an artwork could impede pedestrian pathways, the importance of which varied depending on their goals. Regarding Probe~C, an open fountain in a shopping center~(\autoref{fig:probe3}), P3 asked: \textit{“is there something around it that could stop me from getting into where the water shoots, or where the water is going to fall? [...] [If] I’m going to have lunch, I don’t want to be all wet.”}

To access artwork descriptions, most participants envisioned using a smartphone application for greater flexibility, but some also suggested physical on-site triggers such as QR codes and buttons that would play descriptions through speakers. While these strategies enabled access without additional devices, improving financial accessibility, participants expressed concerns about the robustness of hardware outdoors, and acknowledged that such solutions might be better-suited for indoor museum or gallery environments as \textit{“the electronics are more protected inside”} (P14).

\subsubsection{AI-Based Description Reliability}
Drawing on their prior experiences and our design exploration, all 16 participants were at least somewhat positive about using AI to describe artwork, emphasizing its potential for more independent and rapid exploration compared to relying on sighted companions. A few viewed AI as \textit{“the best thing that ever came out for the blindness community”} (P1), while others were more measured. Some already used AI as a first step to interpreting artwork: \textit{“if I can snap a picture and get an AI description, [...] that’s always a good start”} (P9). 

\input{tables/questions}

During the design exploration, all participants asked follow-up questions after hearing the initial descriptions, with ten mentioning they required additional detail to form an adequate understanding of the artwork. Participants asked 98 compound questions ($M = 6.1$, $SD = 2.5$) consisting of 152 discrete sub-questions. For example, a compound question such as \textit{“Can you describe the colors, how the art is composed, and what techniques are used?”} had three sub-questions about color, composition, and art style. Most commonly, participants asked about artwork subjects or content, cultural or historical significance, and composition~(\autoref{table:questions}), and nine questions simply asked for more information~(e.g.,~\textit{“tell me more”}). For the design probe descriptions, P8 shared that \textit{“the AI did pretty good overall and then I was able to ask follow-up questions for things I was unsure of.”}  P7 visually engaged with the probes and thought the short description helped to interpret Probe~B~(\autoref{fig:probe2}), but wanted to query more about details she missed.

Of the 98 AI-generated responses to follow-up questions, five featured inaccuracies and 13 featured refusals due to incomplete context. Inaccuracies included misdescribing composition details, mischaracterizing the art type, or reinforcing existing content errors~(e.g.,~\textit{“The central white rocket or spire-shaped structure appears to represent a skyscraper”} for Probe~E). Refusals centered on artist information, cultural details, or artwork location~(e.g.,~\textit{“Without specific location information [...] I can't identify the exact city or shopping district”} for Probe~C). While AI was useful for generating baseline descriptions, its variability and lack of precision frustrated some participants as \textit{“it just wasn’t reliable”} (P12).

The importance of accuracy varied by context. While most participants thought accuracy was not as critical when exploring artwork for their own personal interest, P4 emphasized the importance of ensuring accuracy if the artwork was \textit{“a landmark, to meet somebody at a certain place.”} Others were disappointed that seemingly innocuous mistakes could be difficult to catch. Our prepared short descriptions contained errors about artwork type and artwork content~(\autoref{fig:probecloseups}). Participants were more accepting of the artwork type error, but suggested that descriptions should indicate artwork tactility, as otherwise \textit{“I’m not going to know what I’m allowed to touch”} (P14). Conversely, multiple participants were frustrated by the misidentification of the canoe, given its cultural significance, with P16 expressing that the mistake \textit{“cheapens the whole view.”}

Even when participants were mostly positive about AI’s potential for generating urban artwork descriptions, they remained cautious: \textit{“it’s sad because so much of the description just sounded full and reasonable [...] the couple little weird hallucination things just make me question the validity of the whole thing”} (P16). Overall, AI was viewed as a powerful access tool, but participants were wary that mistakes could cause minor or major misconceptions.

\section{Discussion}
Through this study, we investigated BLV people’s preferences for discovering and engaging with public and urban art, finding that they perceived it as more place-based, interactive, and approachable than non-public art. To discover urban artwork, participants appreciated both spontaneous notifications and large-scale repositories. While some existing art accessibility methods, such as descriptions and tactile support~\cite{lizhang2023understanding, cavazos2021accessible, rector2017eyes, holloway2019making}, were transferable to public settings, participants described urban-specific considerations such as safety and environmental integration. Through our design exploration, we also investigated how generative AI could fill a gap in current art access techniques to support urban art experiences. Participants aligned with prior findings about AI’s potential for independent art exploration~(e.g.,~\cite{chheda2024engaging, chheda2025artinsight}), exemplified by their follow-up questions to the AI system, but they also highlighted critical accuracy concerns about cultural and community understanding for urban artwork in particular. Furthermore, our findings demonstrate that urban settings present fundamentally distinct challenges that cannot be addressed solely by applying existing practices: urban art exists within dynamic public spaces where safety is paramount, and public art is often deeply related to community culture. We detail how prior BLV art and urban accessibility efforts can and cannot directly translate to urban art accessibility in the following sections, summarized in~\autoref{table:translation}. Ultimately, we demonstrate the importance of expanding art accessibility beyond institutional settings and public space accessibility research beyond functional needs to include creative and cultural dimensions.

\input{tables/translation}

\subsection{Social and Logistical Considerations for Urban Art Accessibility}
\subsubsection{Multisensory Accessibility Methods in Shared Public Spaces}
Participants expressed strong interest in tactile access to urban art, similar to museums and galleries, but outdoor public spaces presented unique challenges: weather exposure, sanitation, and risks of vandalism or theft. The “Hands to the Wall”~\cite{handstothewall} initiative, launched in 2018 in Santiago, Chile, provided tactile reliefs of six murals, demonstrating a successful yet geographically limited first step toward urban art access. In the absence of official access measures, especially considering limited resources available for public art~\cite{butler2023gallery}, there remain opportunities to enhance multisensory experiences. At an individual level, providing materials for visitors to create their own representations could be helpful, similar to findings from prior work on tactile graphics for recreation~\cite{clepper2025would, phutane2022tactile}. More scalable technology-oriented efforts include AI-generated 3D models to streamline replica creation, platforms for sharing crowdsourced or AI-created tactile representations, and computer vision models to validate submissions against the original artwork. Where physical installations are feasible, we recommend using durable, non-conductive, and easily sanitized materials and tethering these aids to existing infrastructure~(e.g.,~signs explaining the artwork, railings) to address weather and theft concerns. How to effectively discourage or prevent vandalism remains an open question.

Participants also suggested integrating additional senses beyond touch \textemdash{} such as hearing~(e.g.,~cultural music) and smell~(e.g.,~relevant scents) \textemdash{} to improve awareness and understanding, similar to prior work on enhancing immersion~\cite{jiang2023beyond, boucherit2025exploring}. These senses could be engaged via personal devices such as wireless headphones and augmented reality headsets, or through public infrastructure such as speakers and scent diffusers installed next to artwork. However, in contrast to museum settings where all visitors are engaging with artwork, participants were concerned that these access methods could disrupt others in the urban space. Indeed, adding ambient noises and scents can make public spaces more sensorially inaccessible to neurodivergent people~\cite{aspectss, mostafa2020architecture, dotch2023understanding, piedade2025towards}, further demonstrating how urban art settings pose new challenges that are distinct from museum settings. It is imperative to acknowledge competing access needs in public spaces; we encourage future HCI work to design and evaluate personalized, opt-in, and location-based multisensory art experiences to ensure urban access for all.

\subsubsection{Enhancing Opportunities and Experiences for Urban Art Exploration}
Unlike controlled and curated museum environments, urban art exists within active public spaces with safety risks: people must be conscious of traffic, crowds, and other hazards while engaging with art. As a result, participants emphasized that learning about urban art should not interfere with navigation, nor should it cause them to obstruct other pedestrians. Prior work has begun exploring how to minimize cognitive overload from navigational guidance while preserving safety~\cite{mascetti2025navgraph}; future work could design ways to support spontaneous art exploration through navigation instructions while limiting auditory overwhelm. For example, AI-powered systems could dynamically adjust description detail or timing based on detected crowd density, traffic patterns, or user walking pace. At an infrastructural level~\cite{saha2021urban}, we recommend that cities designate art appreciation areas with seating options and shade, similar to bus stops, to allow both BLV and sighted people space and time to safely interact with urban artwork. 

Furthermore, participants felt public art could facilitate shared social experiences and conversations, echoing prior literature on how urban art can serve as an agora (i.e., a gathering place for discussion)~\cite{knight2008public}. Specifically, both commissioned installations and unsanctioned works such as graffiti can amplify community voices and grassroots expression. While unsanctioned artworks may be difficult to discover due to obscure locations and limited coverage, as long as they are visible to sighted people, BLV people should also have access to informal art forms. Future efforts could invite BLV people into the process of creating new urban artworks, and HCI research could develop co-design tools that enable BLV artists to create and preview both visual and multisensory components of their artwork. This ultimately encourages broader participation in the public art scene, a critical advantage of community-oriented urban artwork compared to curated museums and galleries.

\subsection{Design Opportunities for AI-Powered Public Art Access}
\subsubsection{Supporting Artwork Discovery}
Whether through a mobile application or smart glasses, participants highlighted their interest in discovering art serendipitously and in the moment. This aligns with preferences discovered in prior work describing the importance of personalized and in-situ push notifications for window shopping as a leisurely activity~\cite{kamikubo2024we, kaniwa2024chitchatguide}. However, shopping malls often have defined directories with information about all shops and points of interest, providing a valuable information source from which AI-powered systems can generate descriptions. Given that urban artwork is often less centrally organized, identifying public art in situ relies on computer vision models trained to identify artworks in the wild from a live camera feed. This could capture image inputs from various angles and depth information for more accurately approximating artwork scale (as compared to the static images used in our study); however, there may be challenges with identifying artwork in busy urban environments~\cite{lavi202217k, novack2020towards}. Models may require fine-tuning to accurately distinguish public artwork (including graffiti~\cite{hughes2009street, lewisohn2008street}) from advertisements, signage, and other visual urban elements~\cite{baldini2022street}. Beyond computer vision methods, we encourage future work to investigate how to integrate human perspectives when creating a directory of urban art and its locations, similar to how digital mapping applications rely on crowdsourced photos and reviews.

\subsubsection{Culturally Informed and Personalized Descriptions}
Our study reaffirmed prior findings about AI being a powerful tool to augment nonvisual access to artwork via descriptions~(e.g.,~\cite{chheda2024engaging, bennett2024painting}). While the \textit{interpretation}~\cite{butler2023gallery, artbeyondsight} of art in museums and galleries should consider the broader context of its curation, urban art was distinct in that the interaction between public art and its physical and cultural location was critical to its full meaning. Valuable attributes to highlight in public art descriptions include: (1)~cultural components~(e.g.,~motifs, symbolism), (2)~tactility and reachability~(e.g.,~mural above ground level, statue on a pedestal), (3)~urban environmental integration~(e.g.,~underneath stairs, on an electrical box), and (4)~community relevance~(e.g.,~relation to neighborhood identity, reflecting local flora and fauna). These suggestions reinforce the importance of communicating diversity and representation through descriptions, building on findings from prior work on downstream implications of describing race, gender, and disability~\cite{bennett2021s}. In the absence of this information, the salience of urban art as a cultural, social, and political tool may be dampened; worse, any uncaught errors can cause cultural misrepresentation and erasure, with compounding downstream effects.

At present, a vast majority of public artworks lack any description. Future research could explore how to use generative AI for creating culturally informed and personalized descriptions in-situ and at scale. Our study demonstrated that an off-the-shelf model, Claude 3.7 Sonnet, generated relatively accurate descriptions from a single static image; participants found many of the AI-generated descriptions for the artwork probes to be satisfactory. 

However, subtle mistakes \textemdash{} such as the error for Probe~E~(\autoref{fig:probe5}) \textemdash{} led to major misunderstandings and cultural erasure, greatly detracting from art experiences and undermining the community connection that is especially meaningful to urban artwork. To further contextualize and validate descriptions, especially for critical educational or cultural scenarios, future AI interventions could leverage a user’s location and reverse image search to integrate information from online content about the artwork~(e.g.,~artists’ social media posts, local news coverage, art databases) and resources about the broader neighborhood. This grounding in verified sources could reduce hallucinations about cultural elements; however, to ensure the most respectful and accurate descriptions, we strongly recommend that artists proactively write descriptions for future artworks, which can also increase their awareness of the importance of accessibility. 

\subsection{Reflection on the Design Space} \label{disc_designspace}
Here, we reflect on our seven-dimensional design space~(\autoref{table:designdimensions}), which we created to (1)~formalize differences between public and private artwork and (2)~identify diverse probes for our design exploration. We aim to illuminate how this design space can be useful for urban and public art and accessibility research.

Our design space guided probe construction to capture variations in urban art contexts through diverse art attributes and engagement experiences, which differ from museum and gallery experiences. These dimensions helped confirm that our probes reflected different \textit{types}, \textit{scales}, \textit{reachabilities}, \textit{visitor goals}, and \textit{crowd densities}~(\autoref{appendix:proberationale}). In turn, this allowed us to elicit distinct responses to different art scenarios, illuminating challenges associated with translating current art access measures such as the practicality of audio descriptions given the importance of auditory input for safety or the tactility of artwork at larger scales. While three probes featured cultural elements \textemdash{} through style in Probe~B~(\autoref{fig:probe2}), content in Probe~D~(\autoref{fig:probe4}), and both in Probe~E~(\autoref{fig:probe5}) \textemdash{} their varied environments drew different reactions regarding safety and desired detail. These diverse contexts also surfaced unique findings surrounding AI accuracy and cultural competency. Our sculptural probes, with Probe~C located on the ground at an outdoor shopping center~(\autoref{fig:probe3}) and Probe~F suspended above a subway platform~(\autoref{fig:probe6}), allowed us to concretely discuss tactile improvements and the placement of tactile aids within urban spaces. Future work could leverage these dimensions to select diverse artwork probes, categorize urban artworks by attributes, or design more varied public art scenarios.

We also noted areas for improvement. First, while we had even coverage of all three artwork \textit{types} (murals, mosaics, and sculptures) in our design space, our study demonstrated how certain urban art types especially appealed to BLV people. For example, we considered fountains such as Probe~C~(\autoref{fig:probe3}) to be sculptures. However, given their intrinsic multisensory interactions, fountains and other interactive, audible, and tactile artworks could constitute a separate art type altogether. Second, though \textit{visitor goal} and \textit{density} can be somewhat proxied by location or purpose of the area~(e.g.,~tourist attraction vs. sidewalk vs. transit center), technology could operationalize these dimensions. For example, real-time crowd density could be estimated through aggregated mobile location data, and visitor goal could be inferred by analyzing nearby points of interest or a user’s trajectory throughout the day. Last, we encountered challenges with addressing \textit{artist purpose} and \textit{sociability} through our probes. Regardless, some participants inferred artwork purpose, and multiple addressed how the often social nature of urban art exploration could enhance their experience and alleviate some safety concerns. Future work could involve artists to uncover their intent or survey participants about typical solo or group art experiences to better capture these dimensions.

\subsection{Limitations and Future Work}
Our study has a few limitations. We conducted virtual interviews with a design exploration to concretely ideate based on diverse urban art probes, AI descriptions, and AI chat interactions. While this enabled us to reach BLV people across the United States with varied public art experiences, we could not evaluate any specific sensory feedback methods~(e.g.,~tactile graphics, musical accompaniments) alongside the descriptions. As physically discovering and engaging with urban art can be a deeply embodied experience, future work should conduct in-situ explorations. Additionally, to avoid participant fatigue, we limited the design exploration to six probes covering diverse aspects of our design dimensions. Given our qualitative methodology, we did not counterbalance probe order; we instead maintained a consistent order that spread the two description errors across different points in the study session. Future work could select a random set of probes from a larger set of artworks to examine how urban art preferences may vary across a wider range of contexts. Lastly, while we designed our prompt to generate AI descriptions based on best practices~(e.g.,~\cite{artbeyondsight, axel2003art, lizhang2023understanding, doore2024images}), we intentionally left the prompt somewhat short and did not adjust the output to encourage organic conversations about inaccuracies and preferences for public art descriptions. Future work could develop multiple detailed prompts to capture objective and subjective details or engage artists in crafting descriptions.

\section{Conclusion}
Given the cultural, social, political, and aesthetic significance of urban art, accessibility is critical. In this paper, we present novel insights on blind and low vision people’s perspectives on public art accessibility, including how technology can better support urban art experiences. We found that art discovery can be facilitated through spontaneous notifications or planned visits, while art engagement can integrate multisensory methods and describe the artwork’s cultural and local significance. We also contribute a design space for future urban and public art access solutions and reflect upon its utility in our study, contrast public and private art environments, and provide design recommendations to address critical access differences between these two settings. Lastly, we investigated how generative AI could support or detract from urban art experiences, finding that BLV people were positive about using AI but current models lacked the cultural context needed for public artwork. From these empirical insights, we synthesize how prior art and urban access work can translate to public art settings and what additional considerations are required. Our work takes a step toward expanding the space of public space accessibility beyond navigation to include creative and cultural dimensions, presenting novel opportunities for future HCI and accessibility work.

\begin{acks}
This work was supported by Apple Inc. Any views, opinions, findings, and conclusions or recommendations expressed in this material are those of the author(s) and should not be interpreted as reflecting the views, policies or position, either expressed or implied, of Apple Inc.

Leah Findlater is also employed by and has a conflict of interest with Apple Inc. This work was conducted independently of any work at Apple Inc.
\end{acks}



%% file: tables/designdimensions.tex
\begin{table*}[h!]
	\renewcommand{\arraystretch}{1.15}
    \begin{center}
    \caption{Our seven design dimensions and examples of what they entail, which assisted in selecting the urban art probes.}
    \Description{Seven dimensions, with four under Art Attributes and three under Engagement Experience. The Art Attribute dimensions are: type, scale, reachability, and artist purpose. The Engagement Experience dimensions are: visitor goal, density, and sociability.}
    \label{table:designdimensions}
        \begin{tabular}{>{\raggedright}p{0.48\textwidth} | >{\raggedright\arraybackslash}p{0.48\textwidth}}
            \toprule
            \textbf{Art Attributes} & \textbf{Engagement Experience} \\
            \midrule[\heavyrulewidth]
            \textbf{Type}: murals, mosaics, sculptures & \textbf{Visitor goal}: cultural, leisure, commuting, etc. \\
            \textbf{Scale}: small $\xleftrightarrow{}$ large & \textbf{Crowd density}: isolated $\xleftrightarrow{}$ crowded \\
            \textbf{Tactility}: fully tactile $\xleftrightarrow{}$ not tactile or out of reach & \textbf{Sociability}: engaging with art by self or with others\\
            \textbf{Artist purpose}: monument, amenity, agora, etc. & \\
            \bottomrule
        \end{tabular}
    \end{center}
\end{table*}

%% file: tables/participants.tex
\begin{table*}[ht!]
	\renewcommand{\arraystretch}{1.15}
    \begin{center}
    \caption{Participant demographics, such as BLV identity, vision details, urban art engagement, and state of residence.}
    \Description{Study participant demographics for 16 participants. Columns from left to right: Participant ID, BLV Identity, Vision Details, Engagement, and State.}
    \label{table:participants}
        \begin{tabular}{>{\raggedright\arraybackslash}p{0.03\textwidth} >{\raggedright\arraybackslash}p{0.135\textwidth} >{\raggedright\arraybackslash}p{0.586\textwidth} >{\raggedright\arraybackslash}p{0.11\textwidth} >{\raggedright\arraybackslash}p{0.04\textwidth}}
            \toprule
            \textbf{ID} & \textbf{BLV Identity} & \textbf{Vision Details} & \textbf{Engagement} & \textbf{State} \\
            \midrule[\heavyrulewidth]
            P1 & Legally blind & Light perception in left eye & Sometimes & NJ \\
            \rowcolor{gray!15}
            P2 & Blind & Born completely blind & Often & WA \\
            P3 & Blind & Born sighted, lost vision 14 years ago & Infrequently & CA \\
            \rowcolor{gray!15}
            P4 & Blind & Born completely blind & Sometimes & FL \\
            P5 & Blind & Can see colors, vision is mostly in small field in left eye & Daily & CA \\
            \rowcolor{gray!15}            
            P6 & Blind & Born completely blind & Often & CO \\
            P7 & Visually impaired & Retinis pigmentosa, limited peripheral vision and depth perception & Daily & KS \\
            \rowcolor{gray!15}
            P8 & Blind & No light perception, lost all vision at 9 & Infrequently & IN \\
            P9 & Blind & Legally blind since birth, completely blind for 15 years & Infrequently & MI \\
            \rowcolor{gray!15}
            P10 & Blind & Born completely blind, “severely hearing impaired” & Infrequently & TX \\
            P11 & Blind & Born sighted, lost vision during life & Sometimes & NJ \\
            \rowcolor{gray!15}
            P12 & Visually impaired & Born sighted, lost vision 9 years ago, double vision in left eye, limited peripheral vision in right eye & Daily & IL \\
            P13 & Legally blind & Less than 10° visual field in left eye, completely blind in right eye & Infrequently & CO \\
            \rowcolor{gray!15}
            P14 & Blind & Progressively lost sight, has light perception & Daily & IN \\
            P15 & Blind & Born with some sight, lost light perception 6-7 years ago & Infrequently & WA \\
            \rowcolor{gray!15}
            P16 & Blind & No vision in right eye, slight peripheral vision in left eye, more vision loss at 30 & Monthly & WA \\
            \bottomrule
        \end{tabular}
    \end{center}
\end{table*}

%% file: tables/questions.tex
\begin{table*}[!hb]
    \vspace{-0.2em}
	\renewcommand{\arraystretch}{1.15}
    \begin{center}
    \caption{Types of follow-up questions asked about the design probes that were present in over 5\% of 152 discrete sub-questions.}
    \Description{Ten of the most common follow-up question types. Columns from left to right: Question Type, Count, Percentage, and Example Question.}
    \label{table:questions}
        \begin{tabular}{>{\raggedright}p{0.251\textwidth}>{\raggedright\arraybackslash}p{0.02\textwidth} >{\raggedright\arraybackslash}p{0.03\textwidth} >{\raggedright\arraybackslash}p{0.62\textwidth}}
            \toprule
            \textbf{Question Type} & \textbf{\#} & \textbf{\%} & \textbf{Example Question} \\
            \midrule[\heavyrulewidth]
            Subject / content & 21 & 13.8 & \textit{“What are the city elements [from the short description]?”} (P5, E) \\
            \rowcolor{gray!15}
            Cultural / historical significance & 20 & 13.2 & \textit{“If it’s a real plane, what was it used for before?”} (P3, F) \\
            Text & 20 & 13.2 & \textit{“Are there a lot of Chinese letters in this? If so, what do they say?”} (P3, D) \\
            \rowcolor{gray!15}
            Composition & 14 & 9.2 & \textit{“Where are the table and figures positioned in relationship to the background?”} (P8, D) \\
            Scale & 13 & 8.6 & \textit{“What are the dimensions of this artwork?”} (P6, B) \\
            \rowcolor{gray!15}
            General question for more detail & 9 & 5.9 & \textit{“Could you give me more details?”} (P8, A) \\
            Boundaries of art in urban space & 9 & 5.9 & \textit{“Is there something preventing people from walking into it?”} (P16, C) \\
            \rowcolor{gray!15}
            Texture / tactility & 9 & 5.9 & \textit{“Is it raised or flat? Does it have any three-dimensional components?”} (P10, A) \\
            Clarifying question & 8 & 5.3 & \textit{“When it says tropical landscape, what do you mean?”} (P2, A) \\
            \rowcolor{gray!15}
            Colors & 8 & 5.3 & \textit{“Are the rocks multicolored?”} (P10, C) \\
            \bottomrule
        \end{tabular}
    \end{center}
\end{table*}

%% file: tables/translation.tex
\begin{table*}[h!]
	\renewcommand{\arraystretch}{1.15}
    \begin{center}
    \caption{Summary of transferable techniques and technology from prior HCI research on art and urban access, as well as considerations unique to urban art surfaced by our study. Here, we provide some key references for each technique in the table: auditory access \cite{rector2017eyes, cavazos2021multi}, tactile access \cite{neumuller20143d, cavazos2021accessible, cho2021tactile}, AI-generated descriptions \cite{doore2024images, chheda2025artinsight}, exploration and awareness \cite{holloway2019making}, navigation between POIs \cite{microsoftsoundscape}, descriptions of POIs \cite{gleason2018footnotes}, and supporting leisure \cite{kamikubo2024we, gupta2024sonicvista}.}
    \Description{Columns from left to right: Prior Work, Transferable Elements, and Additional Considerations for Urban Art. Prior work on art access: auditory access, tactile access, AI-generated descriptions, and exploration and awareness. Prior work on urban access: navigation between POIs, descriptions of POIs, and supporting leisure.}
    \label{table:translation}
        \begin{tabular}{>{\raggedright}p{0.02\textwidth} | >{\raggedright}p{0.22\textwidth} | >{\raggedright}p{0.24\textwidth} | >{\raggedright\arraybackslash}p{0.44\textwidth}}
            \toprule
              & \textbf{Prior Work} & \textbf{Transferable Techniques} & \textbf{Unique Considerations for Urban Art} \\
            \midrule[\heavyrulewidth]
            \parbox[t]{2mm}{\multirow{7.5}{*}{\rotatebox[origin=c]{90}{
            Art Access}}} & Auditory access & Musical accompaniments, earcons / sound effects & Avoid disruption to others since not everyone will want to engage with urban artwork \\
             & Tactile access & 3D models, 2.5D reliefs, 2D Braille labels or tactile graphics & Design with materials in mind, considering weather exposure, sanitation, and vandalism \\
             & AI-generated descriptions & Contextual descriptions, interactive follow-up & Provide cultural nuance and connection between the art and its environment \\
             & Exploration and awareness & Online repositories to pre-plan artwork visits & Support spontaneity while also prioritizing safety in the urban space \\
            \midrule
            \parbox[t]{2mm}{\multirow{5.5}{*}{\rotatebox[origin=c]{90}{Urban Access}}}  & Navigation between POIs & Spatial audio to localize POIs in the urban space & Integrate POIs beyond those which are solely functional (e.g.,~obstacles, crosswalks) \\
             & Descriptions of POIs & Functional and visual descriptions & Provide detailed descriptions grounded in BLV people’s artistic preferences \\
             & Supporting leisure & Push / pull notifications, sonification & Extend beyond shopping and physical activity to include artistic and cultural engagement \\
            \bottomrule
        \end{tabular}
    \end{center}
\end{table*}

%% file: tables/proberationale.tex
\section{Design Exploration - Artwork Probe Rationale} \label{appendix:proberationale}
\begin{table*}[hbt!]
    \renewcommand{\arraystretch}{1.15}
    \begin{center}
    \caption{Attributes of each of the six artwork probes used during the design exploration.}
    \Description{Attributes for the artwork probes in their order of presentation. Columns from left to right: Probe, Location, Type, Scale, Reachability, Goal, and Density. Small images of the probes are also included for reference, and are briefly described here. Probe A: A fantastical aquatic mural on an electrical box located at an intersection. Probe B: A floral mosaic on a wall beneath a staircase at a tourist attraction. Probe C: An open fountain with mini animal sculptures in an outdoor shopping center. Probe D: A pan-Asian cultural mural along the wall of a building above ground level. Probe E: A mosaic with Indigenous American elements along a staircase at a transit station. Probe F: A large sculpture suspended within an open atrium space in a transit station.}
    \label{table:proberationale}
        \begin{tabular}{>{\raggedright}p{0.30\textwidth} | >{\raggedright\arraybackslash}p{0.125\textwidth} | >{\raggedright}p{0.078\textwidth} | >{\raggedright\arraybackslash}p{0.07\textwidth} | >{\raggedright\arraybackslash}p{0.12\textwidth} | >{\raggedright\arraybackslash}p{0.085\textwidth} | >{\raggedright\arraybackslash}p{0.08\textwidth}}
        \toprule
        \textbf{Probe} & \textbf{Location} & \textbf{Type}& \textbf{Scale} & \textbf{Reachability} & \textbf{Goal} & \textbf{Density} \\
        \midrule[\heavyrulewidth]
        (a) Untitled - Rhodora Jacob \\ \vspace{0.2em}
        \includegraphics[width=0.17\textwidth]{figures/1-mural.png} & 
            On the sidewalk at an intersection & 
            Mural & 
            Small & 
            Reachable, but not interpretable via touch & 
            Walking by & 
            Somewhat crowded \\
        \midrule
        (b) Northwest Microcosm - Clare Dohna \\ \vspace{0.2em}
        \includegraphics[width=0.17\textwidth]{figures/2-mosaic.png} & 
            Beneath a staircase at a tourist attraction & 
            Mosaic & 
            Medium & 
            Mostly reachable, except for the top left corner & 
            Seeking a cultural experience & 
            Generally very crowded \\
        \midrule
        (c) Water Frolic - Georgia Gerber \\ \vspace{0.2em}
        \includegraphics[width=0.17\textwidth]{figures/3-sculpture.png} & 
            On the ground at an outdoor shopping center & 
            Sculpture & 
            Small & 
            Reachable, but near the ground and spraying water & 
            Shopping and socializing & 
            Somewhat crowded \\
        \midrule
        (d) Untitled - Lauren YS \\ \vspace{0.2em}
        \includegraphics[width=0.17\textwidth]{figures/4-mural.png} & 
            Along a building wall above ground level in a cultural district & 
            Mural & 
            Large & 
            Not reachable, given its height above the ground & 
            Walking by, or seeking a cultural experience & 
            Not very crowded \\
        \midrule
        (e) Untitled - Laura Sindell \\ \vspace{0.2em}
        \includegraphics[width=0.17\textwidth]{figures/5-mosaic.png} & 
            Along the staircase of a transit station & 
            Mosaic & 
            Large & 
            Partially reachable, except for the top sections & 
            Passing through & 
            Could be crowded during rush hour \\
        \midrule
        (f) Jet Kiss - Mike Ross \\ \vspace{0.2em}
        \includegraphics[width=0.17\textwidth]{figures/6-sculpture.png} & 
            Suspended within an open atrium space in a transit station & 
            Sculpture & 
            Large & 
            Not reachable, given its height above the ground & 
            Passing through & 
            Could be crowded during rush hour \\
        \bottomrule
        \end{tabular}
    \end{center}
\end{table*}

%% file: tables/probedescriptions.tex
\section{Design Exploration - Artwork Probe Descriptions} \label{appendix:probedescriptions}
\begin{table*}[hbt!]
    \renewcommand{\arraystretch}{1.15}
    \begin{center}
    \caption{Short and long descriptions presented to participants during the design exploration portion of the interviews.}
    \Description{Descriptions for the artwork probes in their order of presentation. Columns from left to right: Probe, Short Description, and Long Description.}
    \label{table:probedescriptions}
        \begin{tabular}{>{\raggedright}p{0.121\textwidth} | >{\raggedright}p{0.182\textwidth} | >{\raggedright\arraybackslash}p{0.637\textwidth}}
        \toprule
        \textbf{Probe} & \textbf{Short Description} & \textbf{Long Description} \\
        \midrule[\heavyrulewidth]
        (a) Untitled - Rhodora Jacob & Utility box street art featuring a whimsical green sea monster with purple eyes, small character riding atop, and tropical landscape on turquoise background. & This vibrant street art transforms an urban utility box into a fantastical aquatic scene. The medium-sized installation prominently displays a cartoonish green sea monster with large purple eyes, yellow teeth, and a pink tongue emerging from turquoise waters. A small human figure in red clothing rides atop the creature's mossy head. The composition includes stylized waves, tropical islands with yellow-orange palm trees, and additional sea creatures on other sides of the box. The bright color palette of greens, blues, purples, and yellows creates a playful, imaginative scene that enlivens the city sidewalk. \\
        \midrule
        (b) Northwest Microcosm - Clare Dohna & Large-scale mural beneath a stairway featuring vibrant tree of life with colorful mandala-like flowers, birds, and wildlife on a concrete wall. & This striking public mural spans a concrete wall beneath an outdoor staircase. The artwork centers around a stylized tree with gray branches that spread across the entire wall. The tree bursts with dozens of colorful, ornate flower blooms in mandala patterns, using vibrant reds, yellows, blues, purples, and oranges. Small birds and wildlife are interspersed among the branches, while additional flowers and plants grow from the ground below. The folk art-inspired design creates a whimsical, nature-filled scene that transforms the otherwise plain urban space into a lively ecosystem of color and pattern. \\
        \midrule
        (c) Water Frolic - Georgia Gerber & Public fountain sculpture featuring bronze frogs and turtles on granite boulders spraying water in a shopping district plaza. & This playful public art installation consists of a water fountain with bronze sculptures of frogs and turtles positioned atop large, rounded granite boulders. The medium-sized fountain occupies a paved plaza in a retail area. Several bronze frogs sit on the stone surfaces, spraying thin arcs of water between them, while turtles rest on another boulder. The natural-colored stones contrast with the metallic bronze sculptures, creating an inviting urban oasis near storefronts, outdoor seating, and a wooden bench. The water feature adds movement and ambient sound to the commercial streetscape. \\
        \midrule
        (d) Untitled - Lauren YS & Large-scale vibrant mural on building wall depicting Asian figures in traditional clothing around a table, with mythical creatures on deep blue background. & This impressive multi-story mural covers an entire building wall, showcasing a gathering of Asian figures in elaborate traditional clothing. The artwork features a striking color palette of deep blues, bright pinks, and rich golds. Central to the composition are people of different ages sharing a meal, surrounded by ornately dressed individuals in ceremonial attire. Mythical elements include dragon motifs, symbolic creatures, and decorative masks. The vivid imagery combines cultural celebration with fantasy elements, creating a dramatic visual narrative that transforms the urban architecture. The artist's signature appears in the upper left corner against the bold royal blue backdrop. \\
        \midrule
        (e) Untitled - Laura Sindell & Large subway tile mosaic depicting stylized city elements with a white rocket shape, yellow totems, and blue background alongside a staircase. & This public art installation features a large-scale ceramic tile mosaic mounted on a brick wall in what appears to be a subway or transit station. The pixelated design includes a prominent white rocket or obelisk shape flanked by two yellow totemic figures with circular tops and patterned bodies. The background uses predominantly deep blue tiles with accents of light blue, green, and other colors creating abstract landscape elements. The mosaic follows a stepped pattern that works with the architectural space, including the staircase visible in the lower right. The colorful geometric style creates a striking contrast against the reddish-brown brick wall surrounding it. \\
        \midrule
        (f) Jet Kiss - Mike Ross & Large-scale suspended installation of pink vintage aircraft parts in an atrium space, with American military insignia visible on fuselage sections. & This massive sculptural installation features deconstructed pink aircraft components suspended throughout a multi-story interior atrium. The artwork includes sections of military plane fuselages, wings, and cylindrical elements, predominantly painted in bright pink with yellow accents. A U.S. military star emblem is visible on one component. The installation hangs dramatically across several floors of what appears to be a public building, with escalators visible below. The industrial metal framework of the building contrasts with the repurposed vintage aircraft parts, creating a striking visual dialogue between the architectural space and the suspended sculpture. \\
        \bottomrule
        \end{tabular}
    \end{center}
\end{table*}